\documentclass{aa}
\usepackage[varg]{txfonts}
\usepackage{natbib}
\usepackage{graphicx}
\usepackage{hyperref}
\usepackage{tablefootnote}

\begin{document}


\title{Bottom-up dust nucleation theory in oxygen-rich evolved stars}

\subtitle{II. Magnesium and calcium aluminate clusters\\}

\author{David Gobrecht\inst{1}
 \and S. Rasoul Hashemi \inst{1}
 \and John M. C. Plane\inst{2}
 \and Stefan T. Bromley\inst{3,4}
 \and Gunnar Nyman\inst{1} 
 \and Leen Decin\inst{5}} 
%
\offprints{D. Gobrecht, \email{dave@gobrecht.ch}}

\institute{Department of Chemistry \& Molecular Biology
 , Gothenburg University,
  Box 462, S-40530 Gothenburg, Sweden
\and School of Chemistry, University of Leeds, Leeds LS2 9JT, United Kingdom
\and Departament de Ci\`encia de Materials i Qu\'imica F\'isica \& Institut de Qu\'imica Te\'orica i Computacional (IQTCUB), Universitat de Barcelona, E-08028 Barcelona, Spain
\and Instituci\'o Catalana de Recerca i Estudis Avan\'cats (ICREA), E-08010 Barcelona, Spain
\and Institute of Astronomy, KU Leuven, B-3000 Leuven, Belgium}

\date{Received -- / Accepted --}

\abstract
{Spinel (MgAl$_{2}$O$_{4}$) and krotite (CaAl$_{2}$O$_{4}$) are alternative candidates to alumina (Al$_2$O$_3$) as primary dust condensates in the atmospheres of oxygen-rich evolved stars. 
Moreover, spinel was proposed as a potential carrier of the circumstellar 13 $\mu$m feature. However, the formation of nucleating spinel clusters is challenging; in particular, the inclusion of Mg constitutes a kinetic bottleneck.}
{We aim to understand the initial steps of cosmic dust formation (i.e. nucleation) in oxygen-rich environments using a quantum-chemical bottom-up approach.}
{Starting with an elemental gas-phase composition, we constructed a detailed chemical-kinetic network that describes the formation and destruction of magnesium-, calcium-, and aluminium-bearing molecules as well as the smallest dust-forming (MgAl$_{2}$O$_{4}$)$_1$ and (CaAl$_{2}$O$_{4}$)$_1$ monomer clusters. 
Different formation scenarios with exothermic pathways were explored, including the alumina (Al$_2$O$_3$) cluster chemistry studied in Paper I of this series.
The resulting extensive network was applied to two model stars, a semi-regular variable and a Mira-type star, 
and to different circumstellar gas trajectories, including a non-pulsating outflow and a pulsating model.
We employed global optimisation techniques to find the most favourable (MgAl$_2$O$_4$)$_n$, (CaAl$_2$O$_4$)$_n$, and mixed (Mg$_x$Ca$_{(1-x)}$Al$_2$O$_4$)$_n$ isomers, with $n$=1$-$7 and x$\in$[0..1], and 
we used 
high level
quantum-chemical methods to determine their potential energies.
The growth of larger clusters with $n$=2$-$7
is described by the temperature-dependent Gibbs free energies. 
} 
{In the considered stellar outflow models, spinel clusters do not form in significant amounts.
However, we find that in the Mira-type non-pulsating model CaAl$_2$O$_3$(OH)$_2$, a hydroxylated form of the calcium aluminate krotite monomer forms at abundances as large as 2$\times$10$^{-8}$ at 3 stellar radii, corresponding to a dust-to-gas mass ratio of 1.5$\times$10$^{-6}$. 
Moreover, we present global minimum (GM) candidates for (MgAl$_2$O$_4$)$_n$ and (CaAl$_2$O$_4$)$_n$, where $n$=1$-$7.
For cluster sizes $n$=3$-$7, we find new, hitherto unreported GM candidates.
All spinel GM candidates found are energetically more favourable
than their corresponding magnesium-rich silicate clusters with an olivine stoichiometry, namely (Mg$_2$SiO$_4$)$_n$. Moreover, calcium aluminate clusters, (CaAl$_2$O$_4$)$_n$, are more favourable than their Mg-rich counterparts; the latter show a gradual enhancement in stability when Mg atoms are substituted step by step with Ca.}
{Alumina clusters with a dust-to-gas mass ratio of the order of 10$^{-4}$ remain the favoured seed particle candidate in our physico-chemical models. However, CaAl$_2$O$_4$ could contribute to stellar dust formation and the mass-loss process. In contrast, the formation of MgAl$_2$O$_4$ is negligible due to the low reactivity of the Mg atom.}
{}
{}

\maketitle 

\section{Introduction}
The importance of kinetics in cosmic dust formation was recently highlighted by \cite{2022FrASS...9.8217T}. In particular, the highly dynamical atmospheres of asymptotic giant branch (AGB) stars, which are affected by convection, pulsational shock waves, and varying light emission, imply too short timescales for chemical equilibrium conditions to prevail \citep{2017A&A...600A.137F}.  
Therefore, it is not surprising that sophisticated chemical-kinetic networks are successful in explaining the abundances  of many molecules, including relevant dust precursors \citep{2016A&A...585A...6G}. 
Chemical equilibrium calculations can also reproduce the observed abundances of the predominant molecules CO, H$_2$O, and SiO in AGB atmospheres with C/O$<$1, but the predictions for
aluminium-, calcium-, and magnesium-bearing oxides, which potentially play a role in dust formation, show discrepancies with observations \citep{Agundez_2020}.

It is well established that silicates represent the 
principal
dust component in oxygen-rich astrophysical environments (see e.g. \citealt{2010LNP...815.....H,2018A&ARv..26....1H}). 
However, kinetic investigations indicate that the formation of silicates, perhaps instigated by the dimerisation  of SiO molecules, is too slow to proceed via free gas-phase molecules \citep{C6CP03629E,10.3389/fspas.2023.1135156}.
Moreover, silicate formation routes instigated by MgO polymerisation are unlikely \citep{1997A&A...320..553K,2019MNRAS.tmp.2040B}.
Instead, it is likely that silicates form on the surfaces of different seed particles.
Owing to the high sublimation temperatures of their condensates, alumina (Al$_2$O$_3$) and titania (TiO$_2$) are prototypical seed particle candidates for silicate formation \citep{2016A&A...585A...6G,2022A&A...668A..35S}.
In the diverse gas-phase mixture of O-rich envelopes, it is 
possible that chemically more complex oxide species play a significant role in circumstellar dust nucleation and seed particle formation.

\
As such, ternary metal oxides, which contain two different metal cations, show a greater structural diversity and are more complex than binary oxides. 
Several ternary oxides that include silicates and titanates are expected to play a significant role in the dust condensation zones of oxygen-rich AGB stars 
\citep{doi:10.1111/j.1365-2966.2011.20255.x,doi:10.1098/rsta.2011.0580,doi:10.1098/rsta.2012.0335}. 
\cite{doi:10.1098/rsta.2012.0335} 
predicted
that CaTiO$_3$ is a kinetically favourable condensation nucleus. In contrast, silicate and titanate clusters that contain Mg show negligible concentrations. 
These results are supported by the fact that so far there has been no unambiguous detection of Mg-bearing molecules in oxygen-rich AGB stars.  
However, circumstellar dust contains a substantial amount of magnesium in the form of Mg-rich silicates \citep{Rietmeijer_1999,doi:10.1111/j.1365-2966.2011.20255.x} and potentially also in the form of 
Mg-containing spinel. Therefore, Mg is assumed to be predominantly in atomic (or ionised) form before it condenses, which is in line with the non-detection of Mg oxides and hydroxides \citep{Decin_2018}.
The situation is similar for CaO and CaOH, of which there has been no detection in circumstellar envelopes.

Recent Atacama Large Millimeter Array (ALMA) observing campaigns carried out by the ATOMIUM\footnote{https://fys.kuleuven.be/ster/research-projects/aerosol/atomium} collaboration addressed the potential molecular precursors of oxygen-rich circumstellar dust, SiO, AlO, AlOH, TiO, and TiO$_2$ \citep{2022A&A...660A..94G}, and their oxidation agents OH and H$_2$O \citep{2023A&A...674A.125B}, 
as well as the specific locations of dust formation in the circumstellar envelopes \citep{2023A&A...671A..96M}. Ca- and Mg-bearing molecules and clusters were observationally not addressed, for the reasons elaborated upon in the previous paragraph, 
but their condensates can be identified in observations of broad spectral dust features and meteoritic stardust analysis.

\citet{2003ApJ...594..483S} investigated the 13 $\mu$m spectral feature that shows an anti-correlation with the silicate features seen at 10 and 18 $\mu$m. Furthermore, they found that stars that show the strongest 13 $\mu$m feature are associated with low to moderate mass-loss rates of $\dot{M}=10^{-8}-1.5 \times 10^{-7}$ M$_{\odot}$yr$^{-1}$. 
The authors conclude that crystalline forms of alumina are likely carriers of the 13 $\mu$m feature. 
Alternatively, \citet{1999A&A...352..609P} suggested spinel as a probable carrier of this dust feature as it shows an additional emission at 16.8 $\mu$m in the laboratory that is also observed in the spectra of some stars. \citet{2001A&A...373.1125F} confirmed the spinel emission at 16.8 $\mu$m and identified a third prominent feature at 32 $\mu$m.

In laboratory studies of meteoritic stardust, two spinel grains with sizes of 230 and 590 nm were identified \citep{Mostefaoui_2004}.
\citet{2010ApJ...717..107G} identified 38 spinel grains, the majority of which  are likely the condensates from red giant and AGB stars.
Later, \citet{2014GeCoA.124..152Z} found 37 individual spinel grains with typical AGB star isotopic anomalies and sizes of 0.8 to 4 $\mu$m. Several of these grains show a blocky appearance, suggesting that they are aggregates of smaller grains.
These studies show clear evidence for the presence of sub-micron-sized stardust grains with a spinel composition.

Calcium-aluminium-rich inclusions are found in carbonaceous chondrite meteorites and are attributed to the first and oldest solids formed in our Solar System \citep{doi:10.1126/science.1226919}. 
Calcium aluminate was found in the crystalline form of the rare mineral krotite in the NWA 1934 meteorite \citep{10.2138am.2011.3693}.

Certain stability limits for different crystalline solids can be predicted from equilibrium vapour pressure measurements. 
In stellar outflows, a condensation sequence with decreasing temperatures and pressures can be derived \citep{2013A&A...555A.119G}. The following sequence is reported:
corundum (Al$_2$O$_3$), gehlenite (Ca$_2$Al$_2$SiO$_7$), spinel (MgAl$_2$O$_4$), forsterite (Mg$_2$SiO$_4$), and enstatite (MgSiO$_3$). 
We note that this sequence relates to the crystalline bulk material and equilibrium conditions and represents a top-down approach.  
In contrast, in this study we model the nucleation of dust seed particles at the (sub-)nanoscale
following a bottom-up approach, which is not restricted to the bulk stoichiometry, sphericity, bimolecular association reactions, or monomeric growth of the nucleating clusters. 

The chemical-kinetic formation routes to alumina dust seed particles were
studied in Paper I of this series \citep{Gobrecht_alumina}.
In the present study, we investigate the possibility of spinel (MgAl$_{2}$O$_4$) and calcium aluminates (CaAl$_2$O$_4$) as seed particles that trigger the onset of dust formation and mass loss in AGB stars.\
Spinel is a ternary oxide that shows the same atomic components as alumina with an additional Mg-O building block. For the calcium aluminate (i.e. krotite), there is an extra Ca-O unit with respect to alumina.
To our knowledge, a theoretical bottom-up investigation of aluminium-bearing ternary oxides in circumstellar envelopes had never before been carried out.
Because of the similarity of spinel and krotite clusters to those of the astrophysically relevant Mg-rich silicate of olivine type, we compare our results to the study of \cite{doi:10.1021/acsearthspacechem.9b00139}.

This paper is organised as follows. 
In Sect. \ref{2} we describe the methods used to derive the structures and kinetic reaction rate coefficients of the molecular and cluster species included in our study.
The results of our investigations, including kinetic modelling, cluster energies and properties, and predictions for larger cluster and dust sizes, are presented in Sect. \ref{3}.
We discuss these results in light of observations and previous studies in Sect. \ref{4}.
Finally, Sect. \ref{5} provides a summary of our findings.

\section{Methods}
\label{2}
\subsection{Chemical kinetics}
\label{2p1}
In the present study we considered a set of reactions that are combined to form a chemical rate network.
The kinetic network entails 
the complete aluminium-oxygen-hydrogen chemistry from \citet{Gobrecht_alumina} 
and 
reactions R26-R37 as well as R48-R57 reported in \cite{Decin_2018} for the magnesium-calcium-oxygen chemistry.
In addition, we included reactions between Al-, Mg-, and Ca-bearing molecules (see Appendix \ref{A1} for details), 
as well as pathways that describe the formation and destruction of spinel and krotite monomers, and derived their 
rate 
coefficients
via state-of-the-art rate theory. 
The corresponding reactions and rate coefficients are summarised in Table \ref{chemnetwork}.
To solve the set of the stiff differential rate equations we used of the Linear Solving of Ordinary Differential Equations (LSODE) solver \citep{2019ascl.soft05021H}.

We employed transition state theory (TST) based methods, which require reaction energetics and rovibrational properties of the reagents and transition states, to derive diagrams of potential energy surfaces (PESs). 
For reactions involving molecular systems containing more than four atoms, rate coefficients were estimated by combining electronic structure calculations with Rice-Ramsperger-Kassel-Marcus (RRKM) statistical rate theory. 
First, the relative energies of the reactants, products and intermediate stationary points on each PES were calculated using the benchmark complete basis set (CBS-QB3) level of theory \citep{doi:10.1063/1.481224} within the Gaussian 16 suite of programs \citep{g16}.

The PESs for seven of these reactions are illustrated in Figs. \ref{pes_john1} and \ref{pes_john2}
(note that the relative energies include vibrational zero-point energy corrections). 
The Master Equation Solver for Multi-Energy well Reactions (MESMER) program \citep{doi:10.1021/jp3051033} was then used to estimate the rate coefficients. 
The methodology is described briefly here (see \citealt{10.1093/mnras/stac1684} for more details). 
The density of states of each stationary point on the PESs was calculated using the vibrational frequencies and rotational constants calculated at the B3LYP/6-311+g(2d,p) level
\citep{1993JChPh..98.1372B,g16}; 
vibrations were treated as harmonic oscillators, and a classical densities of states treatment was used for the rotational modes.

Microcanonical rate coefficients for the dissociation of intermediate adducts, either forwards to products or backwards to reactants, were determined using inverse Laplace transformation to link them directly to the relevant capture rates that were calculated using long-range TST \citep{doi:10.1063/1.1899603}. 
The probability of collisional transfer between discretised bins was estimated using the exponential down model \citep{Gilbert1990TheoryOU}, with the average energy for downward transitions, $<\Delta$E$>_{down}$, set to 200 cm$^{-1}$ with no temperature dependence for collisions with H$_2$. 
The Master Equation was then solved for each reaction to yield rate coefficients for recombination and bimolecular reaction at specified pressures and temperatures; 
it should be noted that at the low densities (n(H$_2$)$<$ 10$^{14}$ cm$^{-3}$) and high temperatures (T$>$1000 K) in a stellar outflow it is only the bimolecular channels that matter. 
The rate coefficients for the reverse reactions were calculated by detailed balance.

\subsection{Cluster candidates}
\label{candidates}
The PES of a cluster containing 
monomer building blocks of seven atoms, 
for example MgAl$_2$O$_4$, is multi-dimensional, intricate, and computationally expensive to obtain. With increasing cluster sizes the number of local minima on the PES grows 
rapidly
and a thorough search 
with electronic structure methods
becomes untractable and prohibitive.
Therefore, an extensive Monte Carlo basin hopping (MC-BH) search is performed \citep{1998cond.mat..3344W} to find favourable candidate isomers for each size and stoichiometry by 
using a simplified PES that is described by the Coulomb-Buckingham pair potential:

\begin{equation}
\centering
U(r_{ij}) =  \frac{q_{i}q_{j}}{r_{ij}} + A_{ij}\exp\left(-\frac{r_{ij}}{B_{ij}}\right) - \frac{C_{ij}}{r_{ij}^6} ,
\label{buck}
\end{equation}

\noindent
where $r_{ij}$ is the relative distance between two ions, $q_i$ and $q_j$ the charges of ions
$i$ and $j$, respectively, and $A_{ij}$, $B_{ij}$ and $C_{ij}$ the Buckingham pair parameters listed in Table \ref{tab1}.
The chosen Coulomb-Buckingham pair parameters correspond to the values of \citet{woodley_ternary}
and we used an in-house modified version of GMIN \citep{PhysRevLett.95.185505}.

\begin{table}
        \caption{Coulomb-Buckingham pair potential parameters used in this study. 
\label{tab1}}    
\begin{tabular}{ c c  l l l l l }
  i & j   &   q$_i$ & q$_j$ & A$_{ij}$ (eV) & B$_{ij}$(\AA{}) & C$_{ij}$ (eV $\rm{\AA}^{-6}$)\\
        \hline
        \rule{0pt}{4ex}    
 Al & O   &   +3 & -2 &  1460.3  & 0.29912 &  0.0  \\
 Mg & O   &   +2 & -2 &  1428.5  & 0.29453 &  0.0  \\
  O & O   &   -2 & -2 & 22764.0  & 0.14900 & 27.88 \\
\end{tabular}
\end{table}
For (CaAl$_2$O$_4$)$_n$ and mixed calcium-magnesium aluminates we did not perform separate
 global optimisation searches. We used the geometries of the lowest-energy candidates of MgAl$_2$O$_4$ clusters to optimise the (CaAl$_2$O$_4$)$_n$, $n$=1$-$7, clusters.
These two metal aluminate species differ only by their alkaline earth metals Mg and Ca. 
Owing to the larger atomic radius of Ca compared to Mg, the calcium aluminate clusters exhibit larger Ca-O bond distances but the overall cluster geometry is largely preserved.

For each cluster stoichiometry investigated, we optimised the 100-200 lowest-energy isomers 
at the quantum level of theory. 
We used the hybrid B3LYP functional 
along with the cc-pVTZ basis set \citep{cc-pVTZ} 
as a compromise between computational cost and accuracy.
As for the kinetic computations described in Sect. \ref{2p1}, the optimisations were performed with the Gaussian16 program suite \citep{g16}.
For the lowest-energy isomers of each cluster stoichiometry and size found, a vibrational frequency analysis is included. The resulting vibrational modes are required to exclude transition states and higher order saddle points, to construct partition functions for the thermodynamic potentials, and to predict the emission spectra of the clusters.



\section{Results}
\label{3}
In this section we present the rate expressions and nucleation routes derived in this study (Sect. \ref{3p1}), the resulting species abundances in the different physico-chemical models (Sect. \ref{3p2}), the characteristics of the most favourable clusters (Sect. \ref{3p3}), and a comparison to their macroscopic (i.e. crystalline) bulk material (Sect. \ref{3p4}). Finally, the vibrational cluster spectra are addressed in Sect. \ref{3p5}.

\subsection{Molecular precursors of aluminate clusters}
\label{3p1}
In addition to the existing rate network for the oxides of aluminium \citep{Gobrecht_alumina}, as well as of magnesium and calcium \citep{Decin_2018}, we included several redox reactions linking aluminium with magnesium and calcium chemistry. 
The prevalent aluminium-bearing molecules are AlO and AlOH, whereas magnesium and calcium are predominantly in atomic form.\
Since the formation of the metal dioxides AlO$_2$, and in particular, MgO$_2$ and CaO$_2$, is hampered by strongly endothermic oxidation reactions, they are not considered in our proposed nucleation schemes. 
The gas-phase reactions between (hydr-)oxides of Al and Mg/Ca, which are included in this study, comprise 
\begin{equation}
\rm{AlO + Mg/Ca \rightleftharpoons MgO/CaO + Al}, 
\label{alo+m}
\end{equation}
\begin{equation}
\rm{AlO + Mg/Ca + M \rightleftharpoons AlOMg/AlOCa + M}, 
\label{term}
\end{equation}
and
\begin{equation}
\rm{AlOH + Mg/Ca  \rightleftharpoons AlOMg/AlOCa + H}.
\label{bim}
\end{equation}

\begin{figure}[h]
\includegraphics[width=0.48\textwidth]{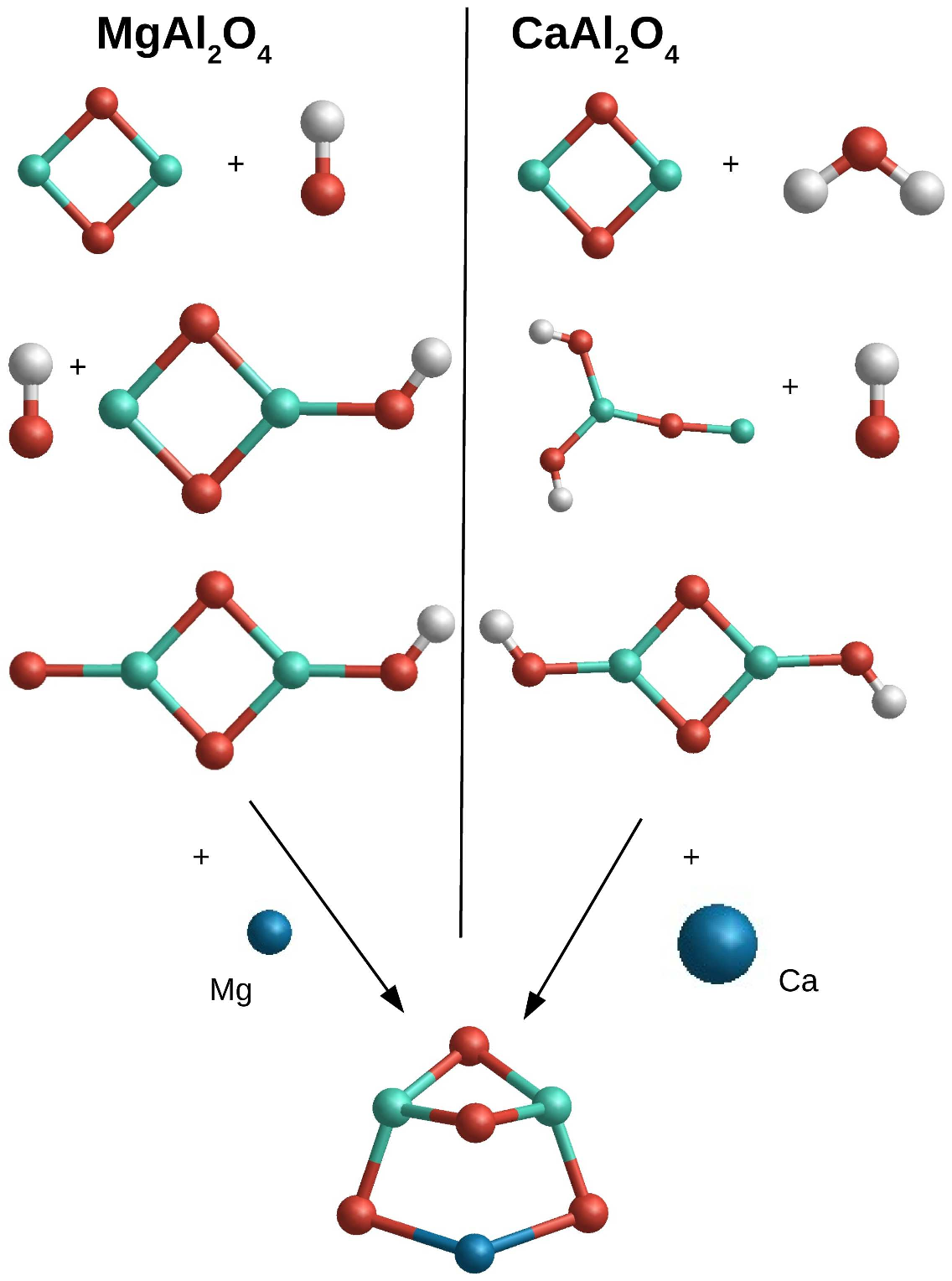}
\caption{Sketch of the exothermic spinel (MgAl$_2$O$_4$) and krotite (CaAl$_2$O$_4$) monomer formation scenarios\label{scenarios}.}
\end{figure}

Owing to the low importance of reactions \ref{alo+m}, \ref{term}, and \ref{bim} to spinel and krotite production, the relevant calculations are described in Appendix \ref{A1}.
To model the spinel and calcium aluminate nucleation processes, we considered different scenarios for the formation of the respective monomer, (MgAl$_2$O$_4$)$_1$ and (CaAl$_2$O$_4$)$_1$.
The most promising formation scenarios are graphically illustrated in Fig. \ref{scenarios}.\\ 

\begin{figure}
\includegraphics[width=0.48\textwidth]{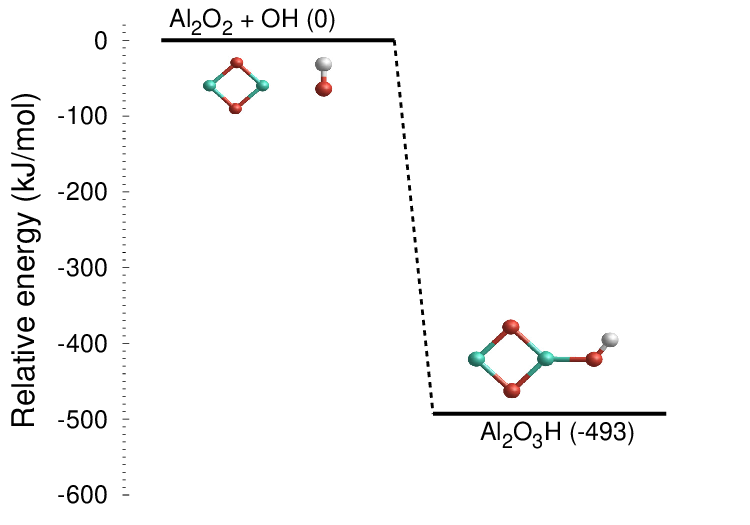}
\includegraphics[width=0.48\textwidth]{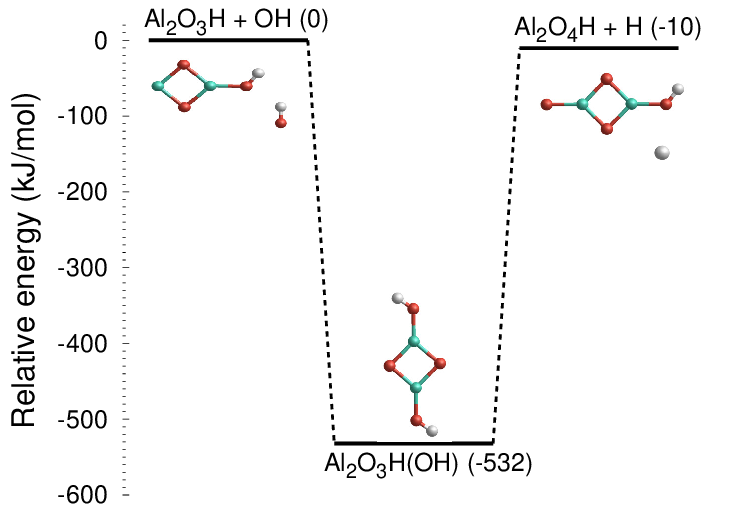}
\includegraphics[width=0.48\textwidth]{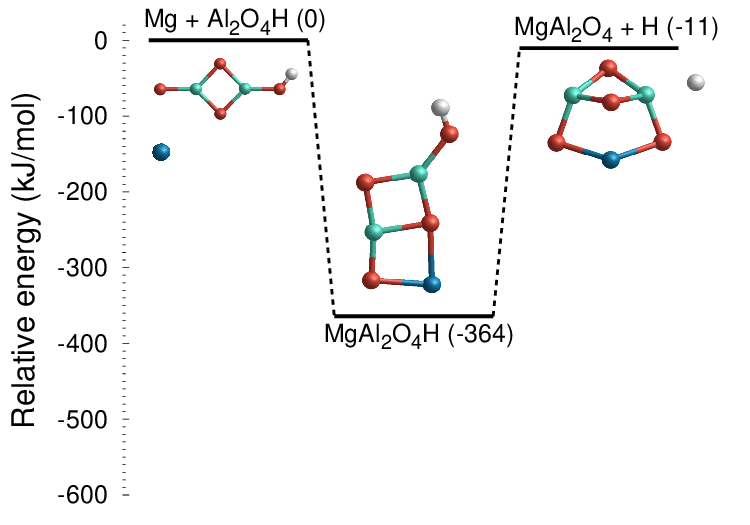}
\caption{Potential energy diagrams for reactions \ref{al2o3hf}, \ref{al2o4jf}, and \ref{mgal2o4f}.\label{pes_john1}}
\end{figure}

As pointed out in Appendix \ref{A1}, the inclusion of Mg and Ca atoms in the first steps of cluster formation is energetically and kinetically inefficient.
Therefore, we investigated the possibility that Mg/Ca atoms are included at a later stage in the formation of MgAl$_2$O$_4$ and CaAl$_2$O$_4$ monomers. 
Starting from Al$_2$O$_2$, which can be readily and exothermically synthesised (see Paper I), a hydrogenated form of the alumina monomer, Al$_2$O$_3$H, can form via the following exothermic termolecular reaction (see also the top-left panel of Fig. \ref{pes_john1}):

\begin{equation}
\rm{Al_2O_2 + OH  + M \rightleftharpoons Al_2O_3H + M}     
\label{al2o3hf}
\end{equation}
with $\Delta_{r}$H(0K)= $-$493 kJ mol$^{-1}$. 

The corresponding rate can be found in \ref{chemnetwork}. 
In principle, reaction \ref{al2o3hf} could also proceed as radiative association, which was not considered in this study, as it proceeds with slow timescales in AGB atmospheres.
Once Al$_2$O$_3$H has formed, it can be oxidised to Al$_2$O$_4$H via

\begin{equation}
\rm{Al_2O_3H + OH \rightleftharpoons Al_2O_4H + H}    
\label{al2o4jf}
\end{equation}
with $\Delta_{r}$H(0K)= $-$10 kJ mol$^{-1}$.
Reaction \ref{al2o4jf} becomes unfavourable for temperatures above $T=$300 K and the reverse process becomes fast, supported by large H abundances. 
If small Al$_2$O$_4$H concentrations persist, the spinel monomer can form via

\begin{equation}
\rm{Mg + Al_2O_4H \rightleftharpoons MgAl_2O_4 + H}
\label{mgal2o4f}
\end{equation}
with $\Delta_{r}$H(0K)= $-$11 kJ mol$^{-1}$.
This reaction 
becomes endoergic ($\Delta_{r}$G(T)$>$0)
above 300 K and the reverse reactions involving atomic H become efficient.\\



\begin{figure}
\includegraphics[width=0.48\textwidth]{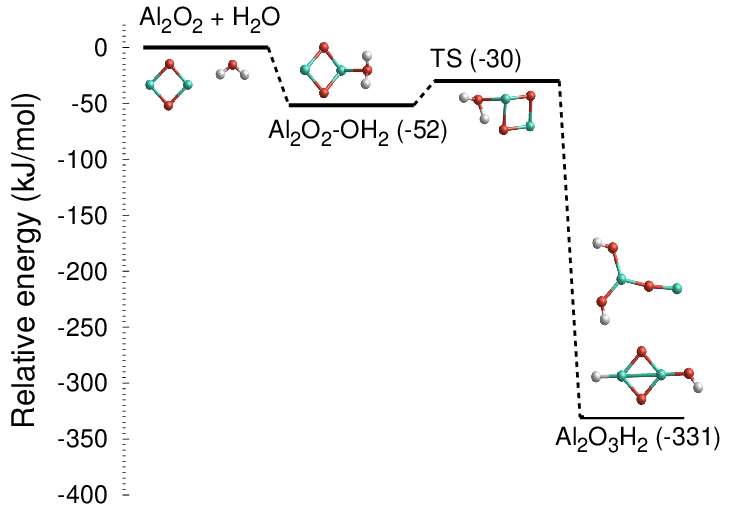}
\includegraphics[width=0.48\textwidth]{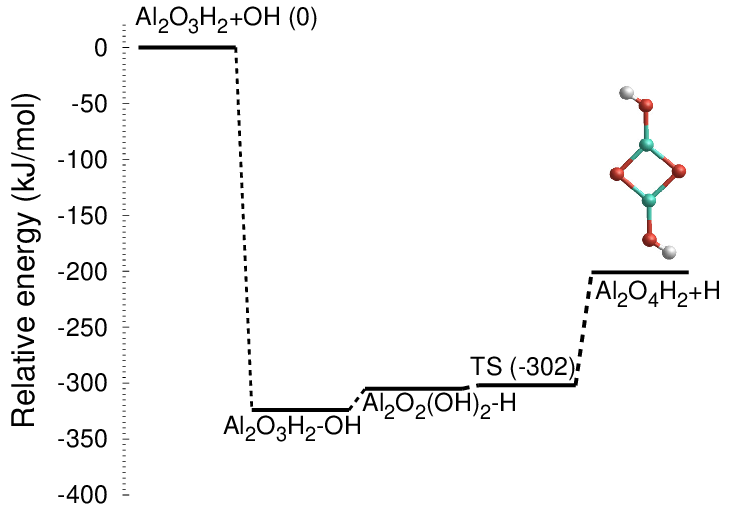}
\includegraphics[width=0.48\textwidth]{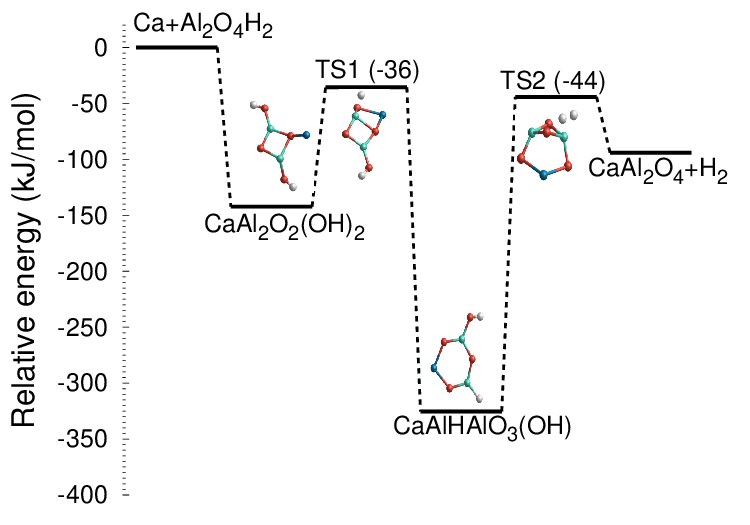}
\includegraphics[width=0.48\textwidth]{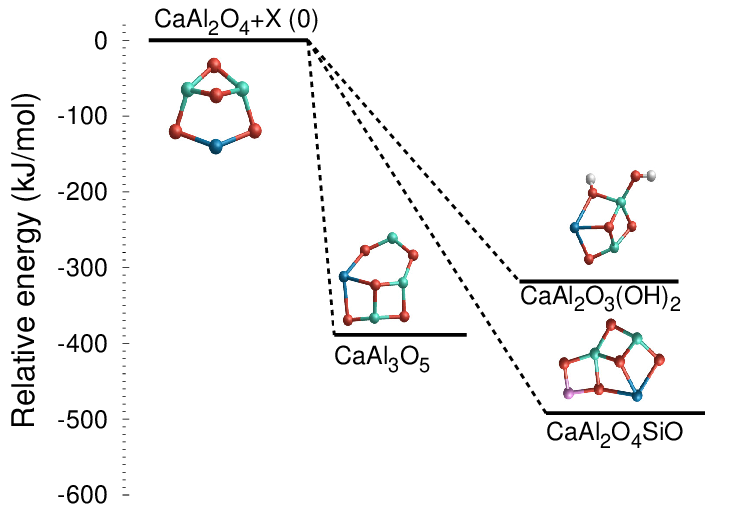}
\caption{Potential energy diagrams for reactions \ref{al2o3h2f}, \ref{al2o4h2f}, \ref{caal2o4f}, and \ref{caal2o5h2} \label{pes_john2}.}
\end{figure}

For the CaAl$_2$O$_4$ monomer formation, we considered a similar starting point as for the spinel formation (see the top-left panel of Fig. \ref{pes_john2}):

\begin{equation}
\rm{Al_2O_2 + H_2O  + M  \rightleftharpoons Al_2O_3H_{2} + M}     
\label{al2o3h2f}
\end{equation}
with $\Delta_{r}$H(0K)= $-$331 kJ mol$^{-1}$, 
where the third body M is assumed to be H$_2$.
However, here H$_2$O is the oxidant leading to a doubly hydrogenated form of the alumina monomer, which exhibits two very favourable isomers with a relative energy of 19 kJ mol$^{-1}$. Al$_2$O$_3$H$_{2}$ can be further oxidised by OH:

\begin{equation}
\rm{Al_2O_3H_2  + OH  \rightleftharpoons Al_2O_4H_{2} + H}     
\label{al2o4h2f}
\end{equation}
with $\Delta_{r}$H(0K)= $-$201 kJ mol$^{-1}$.
In principle, Ca could react exothermically with Al$_2$O$_4$H, which can 
be produced by reaction \ref{al2o4jf}. However, Al$_2$O$_4$H$_2$ forms more quickly due to large H$_2$O concentrations, as compared to OH. Therefore, as a dominant route for making calcium aluminates, atomic Ca can react with Al$_2$O$_4$H$_{2}$:

\begin{equation}
\rm{Ca + Al_2O_4H_2  \rightleftharpoons CaAl_2O_4 + H_{2}}     
\label{caal2o4f}
\end{equation}
with $\Delta_{r}$H(0K)= $-$94 kJ mol$^{-1}$,
leading to the formation of the CaAl$_2$O$_4$ monomer and molecular hydrogen, H$_{2}$.
Despite the reverse reaction being endothermic, it involves H$_2$, which is very abundant in space 
and leads to an effective destruction of CaAl$_2$O$_4$. 
In order to prevent its destruction,
the monomer might further react with abundant oxygen-bearing molecules such as H$_2$O, SiO, or AlO by the exothermic processes

\begin{equation}
\begin{array}{l}
\rm{CaAl_2O_4 + H_2O + M \rightleftharpoons CaAl_2O_3(OH)_2 + M}\\ 
\rm{CaAl_2O_4 + SiO + M \rightleftharpoons CaAl_2O_4SiO + M} \\
\rm{CaAl_2O_4 + AlO + M \rightleftharpoons CaAl_3O_5 + M,}
\end{array}
\label{caal2o5h2}
\end{equation}
with reaction enthalpies of $\Delta_{r}$H(0K)= $-$318 kJ mol$^{-1}$, $\Delta_{r}$H(0K)= $-$392 kJ mol$^{-1}$, and $\Delta_{r}$H(0K)= $-$488 kJ mol$^{-1}$, respectively.
We note that the analogous reaction pathway involving Al$_2$O$_4$H$_2$ and Mg is quite endothermic (+74 kJ mol$^{-1}$). Hence, the spinel monomer cannot be formed in the same manner as CaAl$_2$O$_4$.


\subsection{Numerical modelling in circumstellar envelopes}
\label{3p2}
In this study 
the numerical modelling of the kinetic reaction network is performed under the conditions given by the hydrodynamic models of two model AGB stars, a semi-regular variable (SRV) and a Mira-type variable (MIRA), that were presented in detail in Paper I. 
We summarise the main physical quantities of these models in Table \ref{modelparms}. Furthermore, we differentiated between non-pulsating models and pulsating models. The initial elemental abundances are taken from the FRUITY stellar evolution database \citep{2015ApJS..219...40C} and correspond to the m1p5zsuntdu3 model with a C/O ratio of 0.75. Molecular abundances are reported as number fractions of the total gas and are considered to be significant, if they exceed values of 10$^{-9}$. 

\begin{table}[h]
\caption{Physical quantities of the stellar models SRV and MIRA\label{modelparms}.} 
\begin{tabular}{c   r r r r r r}
quantity     & M$_\star$ & n$_\star$ & T$_\star$  & v$_\infty$ & v$_s$ & P\\
unit         & M$_\odot$ & cm$^{-3}$ & K  & km s$^{-1}$ & km s$^{-1}$ & days \\
\hline
SRV  &  1.2      & 1$\times$ 10$^{14}$ & 2400  & 5.7 & 10 & 332 \\
MIRA &  1.0      & 4$\times$ 10$^{14}$ & 2000  & 17.7 & 20 &  470 \\
\end{tabular}
\end{table}

\subsubsection{Non-pulsating models}
\begin{figure}[h]
\includegraphics[width=0.48\textwidth]{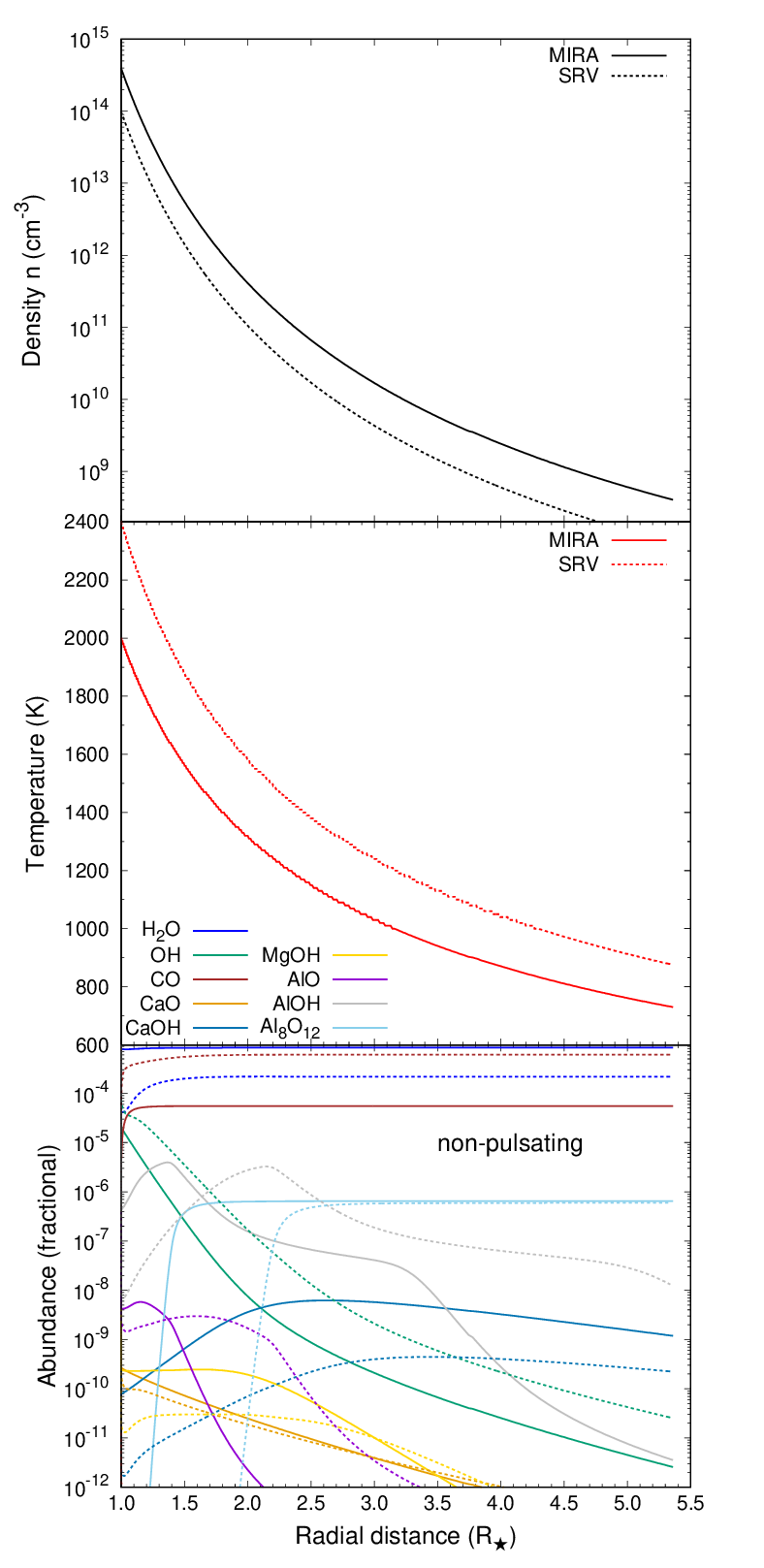}
\caption{Non-pulsating models. \textit{Top panel:} Gas number densities (in cm$^{-3}$) as a function of the radial distance.\
\textit{Middle panel:} Gas temperatures (in K) as a function of the radial distance.\
\textit{Bottom panel:} Fractional abundances of the prevalent gas-phase molecules H$_2$O, OH, and CO and the metal oxides and hydroxides MgO, MgOH, CaO, CaOH, AlO, AlOH, and Al$_8$O$_{12}$ as a function of the radial distance in the non-pulsating models. Solid lines represent the MIRA model and dashed lines the SRV model.\label{nonpMain}}
\end{figure}

The abundances of the prevalent molecules H$_2$O, OH, and CO, together with the metal oxides and hydroxides MgO, MgOH, CaO, CaOH, AlO, and AlOH, including Al$_8$O$_{12}$ as a representative of alumina dust particles, are shown in Fig. \ref{nonpMain} for the non-pulsating outflow model models. 
The species abundances show a very similar behaviour to the models of Paper I of this series.
In the MIRA non-pulsating model, H$_2$O is the dominant O-containing molecule followed by CO, whereas the order is reversed in the SRV non-pulsating outflow, where CO is most abundant. For OH, AlOH, AlO, and Al$_8$O$_{12}$ the two non-pulsating models show similar maximum abundances, but the peaks in the MIRA models occur at smaller radial distances compared with the SRV model. This is a consequence of the lower gas densities and higher temperatures in the SRV model.
CaO exhibits a similar behaviour in both model stars, decreasing from 10$^{-10}$ at the photosphere to lower values farther out. CaOH is one to two orders of magnitude more abundant in the MIRA model compared with the SRV outflow. The most abundant Mg-bearing molecule is MgOH, with abundances below 3$\times$10$^{-10}$ in the entire computational range. 
MgO shows negligible abundances below 10$^{-12}$ in both the SRV and MIRA models.

\begin{figure}[h]
\includegraphics[width=0.48\textwidth]{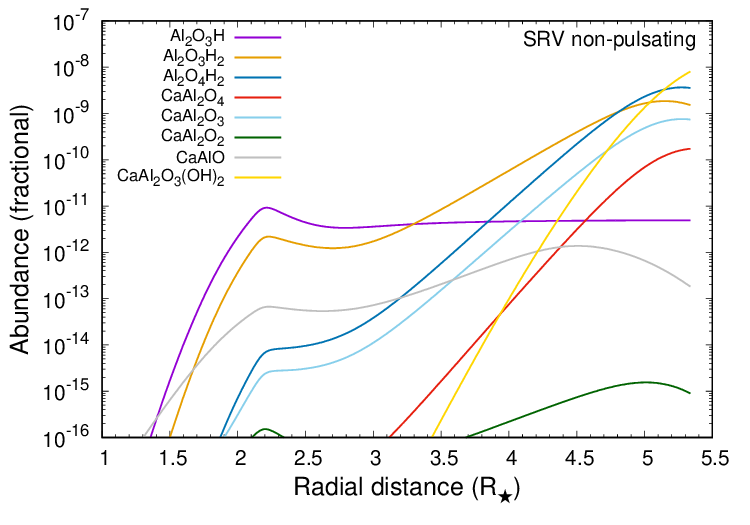}
\caption{Fractional abundances of the molecular precursors related to the aluminate formation and nucleation as a function of the radial distance in the non-pulsating SRV model.\label{srvnonp}}
\end{figure}

In Fig. \ref{srvnonp} the model abundances of the molecular precursors related to the formation of MgAl$_2$O$_4$ and CaAl$_2$O$_4$ are shown as a function of the radial distance for the SRV model. 
At around 2.2 R$_\star$ these species show a distinct peak, 
where Al$_2$O$_3$H and Al$_2$O$_3$H$_2$ obtain abundances in the range of 10$^{-12}-$10$^{-11}$. This peak is related to the emergence of Al$_8$O$_{12}$ clusters as shown in Fig. \ref{nonpMain}.
Whereas the Al$_2$O$_3$H abundance remains approximately constant for larger radial distances and does not lead to MgAl$_2$O$_4$ formation, Al$_2$O$_3$H$_2$ keeps increasing and eventually triggers the formation of Al$_2$O$_4$H$_2$, CaAl$_2$O$_3$, CaAl$_2$O$_4$, and most prominently, CaAl$_2$O$_3$(OH)$_2$. 
In the SRV non-pulsating model, spinel formation is therefore ineffective and the significant calcium aluminate formation occurs only at distances larger than 4 R$_\star$. This indicates that CaAl$_2$O$_4$ is not a primary seed particle.

\begin{figure}[h]
\includegraphics[width=0.48\textwidth]{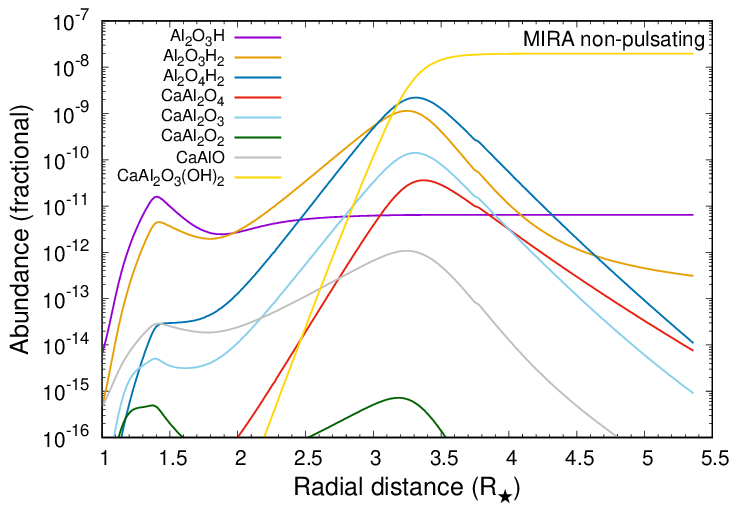}
\caption{Fractional abundances of the molecular precursors related to the aluminate formation and nucleation as a function of the radial distance in the non-pulsating MIRA model.\label{miranonp}}
\end{figure}

In the MIRA non-pulsating model, we find a similar abundance peak of the species Al$_2$O$_3$H and Al$_2$O$_3$H$_2$ as in the SRV non-pulsating model, but at closer distances around 1.4 R$_\star$ (see Fig. \ref{miranonp}). This radial distance marks the onset of the alumina cluster nucleation in the MIRA non-pulsating model (see Fig. \ref{nonpMain}).
This effect was noted already in Paper I of this series on alumina nucleation, and is related to the comparatively higher densities and lower temperatures in the MIRA models. 
Between 2 R$_\star$ and 3 R$_\star$ the species related to CaAl$_2$O$_4$ increase in abundance, leading to a noticeable formation of CaAl$_2$O$_3$(OH)$_2$ for radii larger than 3.5 R $_\star$, although still a factor of $\sim$ 30 lower than the alumina tetramer (Al$_8$O$_{12}$) abundance.

\subsubsection{Pulsating models}
\begin{figure}[h]
\includegraphics[width=0.48\textwidth]{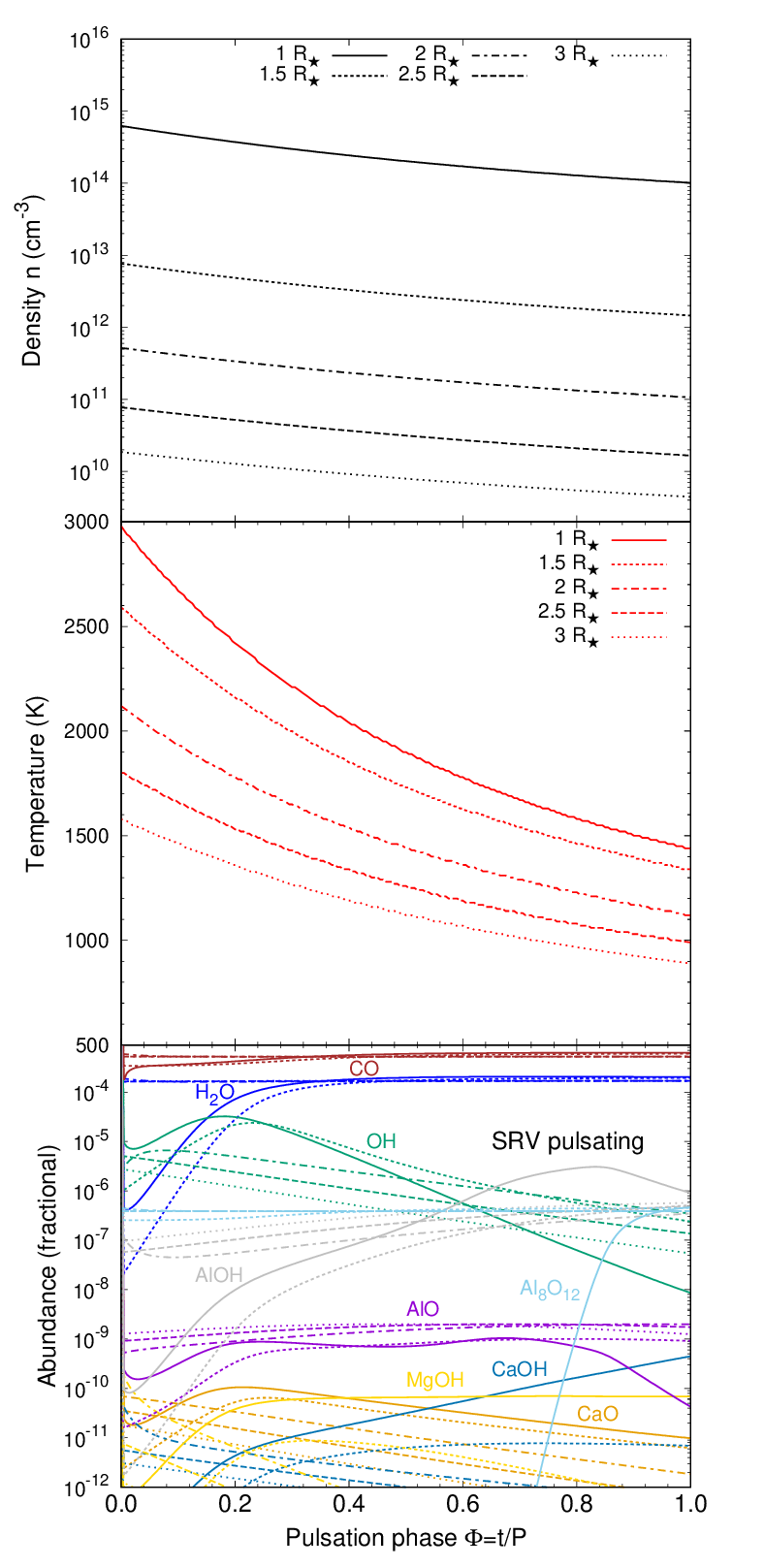}
\caption{Pulsating SRV model.\
\textit{Top panel:} Gas number densities (in cm$^{-3}$) as a function of the pulsation phase, i.e. time, and for the grid of radial distances $r$=1 R$_{\star}$$-$3 R$_{\star}$.\
\textit{Middle panel:} Gas temperatures (in K) as a function of the pulsation phase, i.e. time, and for the grid of radial distances.\
\textit{Bottom panel:} Fractional abundances of the prevalent gas-phase molecules H$_2$O, OH, and CO and metal oxides and hydroxides MgO, MgOH, CaO, CaOH, AlO, AlOH, and Al$_8$O$_{12}$ as a function of the pulsation phase\label{srvpuls}.}
\end{figure}

The output of the SRV pulsating model is shown in Fig. \ref{srvpuls}. The three panels illustrate the total gas number densities, the gas temperature, and the abundances of the prevalent species as a function of pulsational phase $\Phi$ and radial distance $r$, which are modelled from 1 R$_{\star}$ to 3 R$_{\star}$ in steps of 0.5 R$_{\star}$. 
The pulsational phase $\Phi$ is a time coordinate that is normalised to the pulsation period P, which is 332 days for the SRV pulsating model (see Table \ref{modelparms}). In the immediate post-shock gas (i.e. $\Phi=$0.0-0.2), where the temperatures and the gas densities are high, the chemistry is controlled by dissociation reactions, which is particularly pronounced for radial distances close to the star. We note that at the initial conditions at 1 R$_{\star}$ and $\Phi$=0.0 the gas is purely atomic.
At later phases the temperatures drop and
and the molecules recombine in the wake of the still dense postshock gas. Overall, the chemistry is dominated by CO and H$_2$O with similar abundances as in the non-pulsating models, and 
in good agreement 
with 
observations \citep{Decin2010}.
The aluminium content is governed by AlOH and alumina dust, represented by the Al$_8$O$_{12}$ clusters, whereas the AlO abundance is two to three orders of magnitude lower. 
Recent observations of the SRV AGB star R Dor and the Mira-type 
AGB star IK Tau deduce lower AlOH abundances \citep{refId0D}. This might have several reasons, which were discussed in Paper I. 
Here, we note that the AlO/AlOH ratio is very sensitive to the AlOH photolysis rate \citep{Mangan2021}. 
As AGB atmospheres are frequently crossed by pulsational shocks, their radiation field is strongly time dependent, which impacts the AlOH photolysis rate.
In addition, we note that the hydroxides CaOH and MgOH as well as the oxides CaO and MgO with abundances below 10$^{-10}$ play a minor role in the SRV pulsating model.

\begin{figure}[h]
\includegraphics[width=0.48\textwidth]{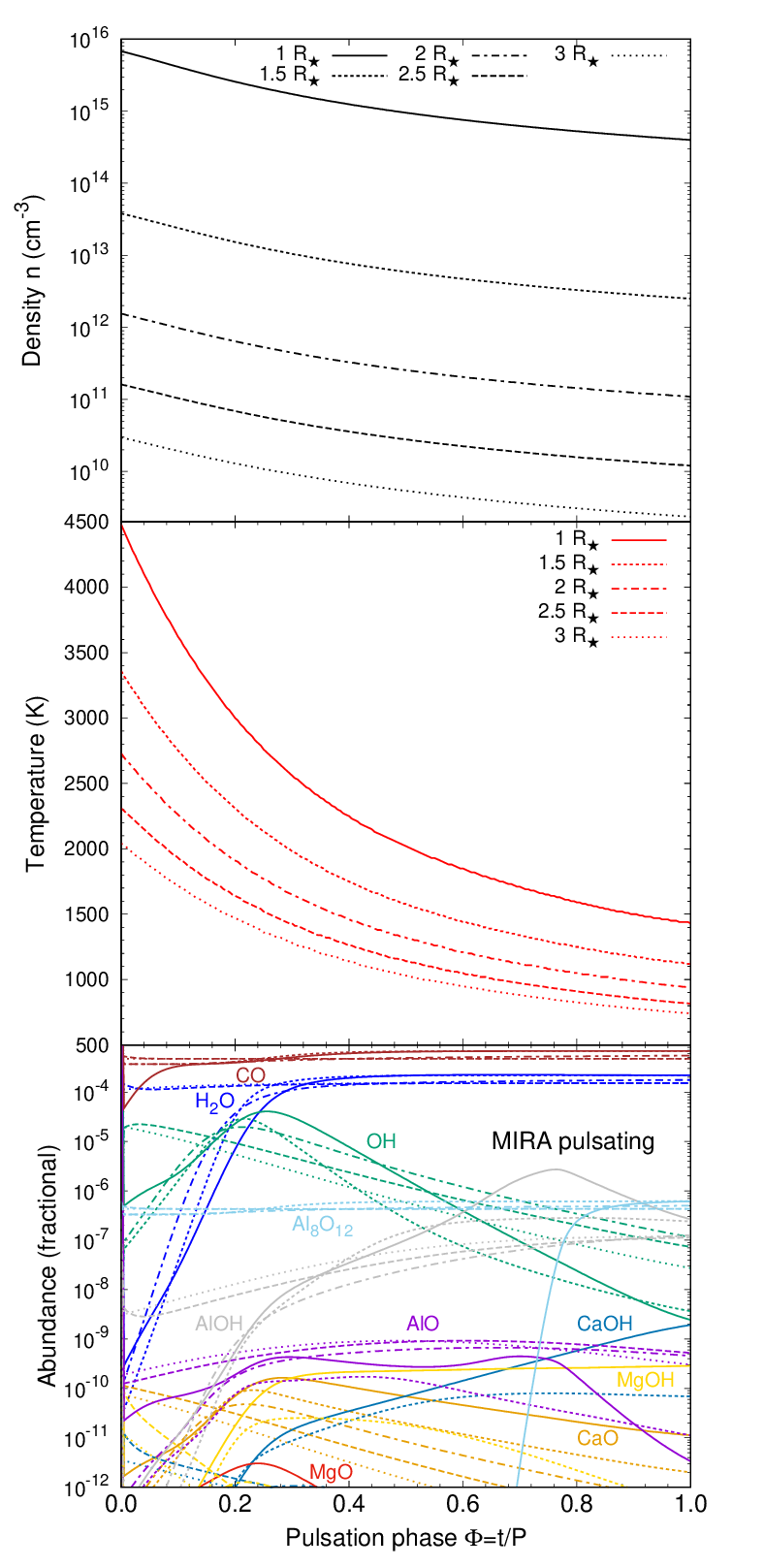}
\caption{Pulsating MIRA model.\
\textit{Top panel:} Gas number densities (in cm$^{-3}$) as a function of the pulsation phase, i.e. time, and the grid of radial distances.\
\textit{Middle panel:} Gas temperatures (in K) as a function of the pulsation phase, i.e. time, and the grid of radial distances.\
\textit{Bottom panel:} Fractional abundances of the prevalent gas-phase molecules H$_2$O, OH, and CO and metal oxides and hydroxides MgO, MgOH, CaO, CaOH, AlO, AlOH, and Al$_8$O$_{12}$ as a function of the pulsation phase and the grid of radial distances\label{mirapuls}.}
\end{figure}

The MIRA pulsating model is presented in Fig. \ref{mirapuls}. Although it shows many similarities with the SRV pulsating model, there are some notable differences. First, the variation in the early postshock gas is larger for most of the species, which is a consequence of the higher shock strength in the MIRA model, as compared with SRV. Second, the hydroxides CaOH and MgOH are more abundant than in the SRV pulsator. Third, a tiny amount of MgO forms at 1 R$_\star$ and phase $\Phi$=0.2.

\begin{figure}[h]
\includegraphics[width=0.48\textwidth]{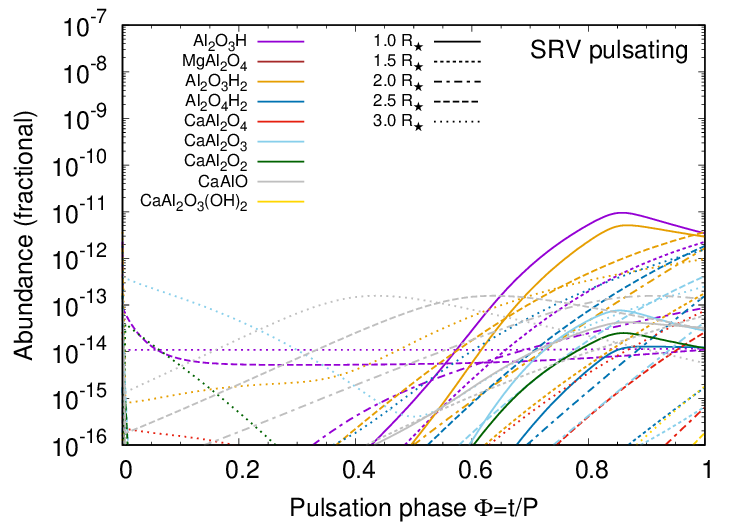}
\caption{Fractional abundances of the considered nucleation clusters as a function of the pulsation phase and the grid of radial distances in the pulsating SRV model.\label{srvpulscluster}}
\end{figure}

By inspecting the abundances of the nucleating species we find that neither CaAl$_2$O$_4$ nor MgAl$_2$O$_4$ forms to any significant extent in the SRV pulsating model (see Fig. \ref{srvpulscluster}). 
The precursors Al$_2$O$_3$H and Al$_2$O$_3$H$_2$ for the MgAl$_2$O$_4$ and CaAl$_2$O$_4$ monomers, respectively, exhibit maximum abundances at 1R$_\star$ and $\Phi$=0.85. However, the subsequent nucleation steps are not efficient and the spinel and calcium aluminate formation does not take place in the pulsating SRV model. 

\begin{figure}[h]
\includegraphics[width=0.48\textwidth]{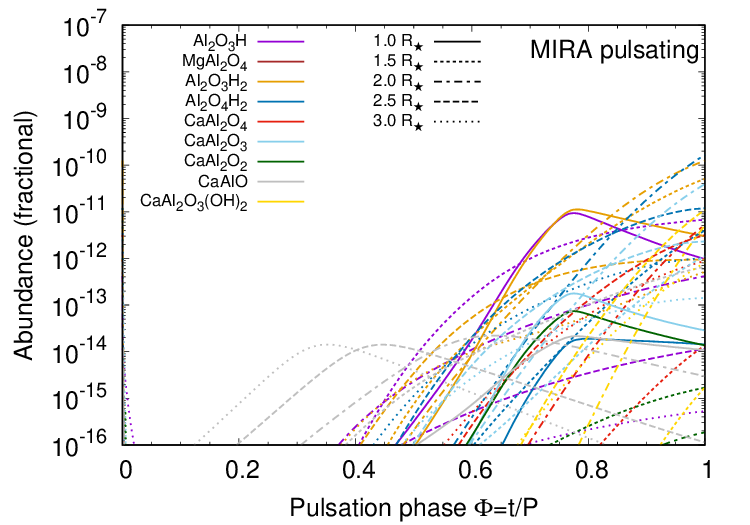}
\caption{Fractional abundances of the considered molecular precursor related to aluminate nucleation as a function of the pulsation phase and the grid of radial distances in the pulsating MIRA model.\label{mirapulscluster}}
\end{figure}

The situation is similar for the MIRA pulsating case, where Al$_2$O$_3$H and Al$_2$O$_3$H$_2$ peak at 1 R$_\star$ and $\Phi$=0.75 (see Fig. \ref{mirapulscluster}). 
In contrast to the SRV pulsating case, some concentrations of Al$_2$O$_3$H$_2$ and Al$_2$O$_4$H$_2$ form at 2 R$_\star$ in the pulsating MIRA model, reaching fractional abundances of 10$^{-10}$ at $\Phi$=1.0. 
These abundances do not persist the passage of the subsequent pulsational shock at 2.5 R$_\star$ and are about four orders of magnitude lower than the solar abundance of Al and Ca. 
Moreover, the Ca inclusion to form the CaAl$_2$O$_4$ monomer and its hydroxylated form is not effective in this model.
Therefore, the contribution of Mg- and Ca-aluminates to circumstellar dust formation is negligible in the MIRA pulsating model.

\begin{figure}[h]
\includegraphics[width=0.48\textwidth]{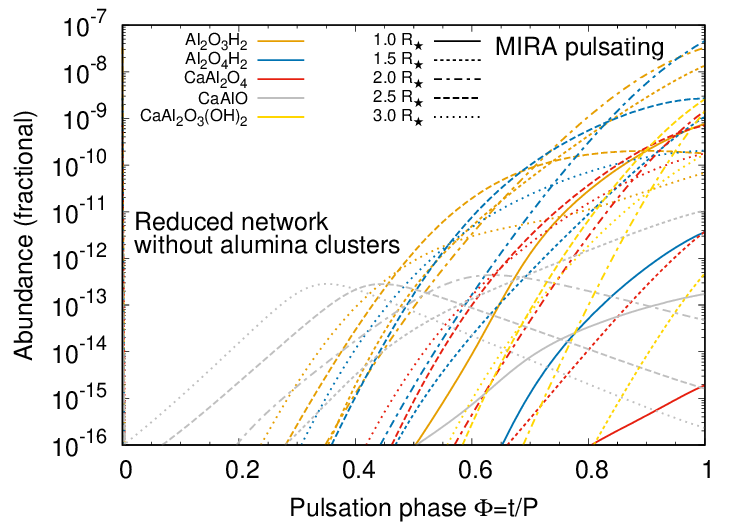}
\caption{Fractional abundances of the considered molecular precursors related to aluminate nucleation as a function of the pulsation phase and the grid of radial distances in the pulsating MIRA model, excluding alumina clustering reactions.\label{mirapulsclusterred}}
\end{figure}

We performed some test calculations by excluding the alumina clustering reactions, that is, reactions involving Al$_x$O$_y$ clusters with $x,y>$2. 
In this reduced chemical network, MgAl$_2$O$_4$ barely forms, but the precursors of 
CaAl$_2$O$_4$ (i.e. Al$_2$O$_3$H$_2$ and Al$_2$O$_4$H$_2$) form in significant amounts of $>$ 10$^{-8}$ at late phases and 2 R$_*$$-$2.5 R$_*$ (see Fig. \ref{mirapulsclusterred}). 
The presence of these precursors leads to a buildup of the CaAl$_2$O$_4$ monomer and its hydroxylated form, showing fractional abundances of $\sim$10$^{-9}$ in the pulsating MIRA model. These amounts are still about two orders of magnitude less than the predictions for alumina clusters, if included. 

\subsection{Spinel (MgAl$_2$O$_4$)$_n$ and krotite (CaAl$_2$O$_4$)$_n$ clusters with sizes $n$=1$-$7}
\label{3p3}
In the following we use the term binding energy, $E_{b}/n$,  defined as the absolute difference between total potential energy of the cluster under consideration and the contribution of its atomic components at $T=$0 K, normalised to the cluster size, $n$. 
This should not be confused with the surface binding energy used in the chemistry of dust grain surfaces and astronomical ices. 
Furthermore, we compared our results for (MgAl$_2$O$_4$)$_n$ with the predictions of \citet{woodley_ternary}, who used an evolutionary algorithm based on interatomic potentials to derive global minimum (GM) configurations.\\

\subsubsection{The monomers ($n$=1)}
\begin{figure}[h]
\includegraphics[width=0.22\textwidth]{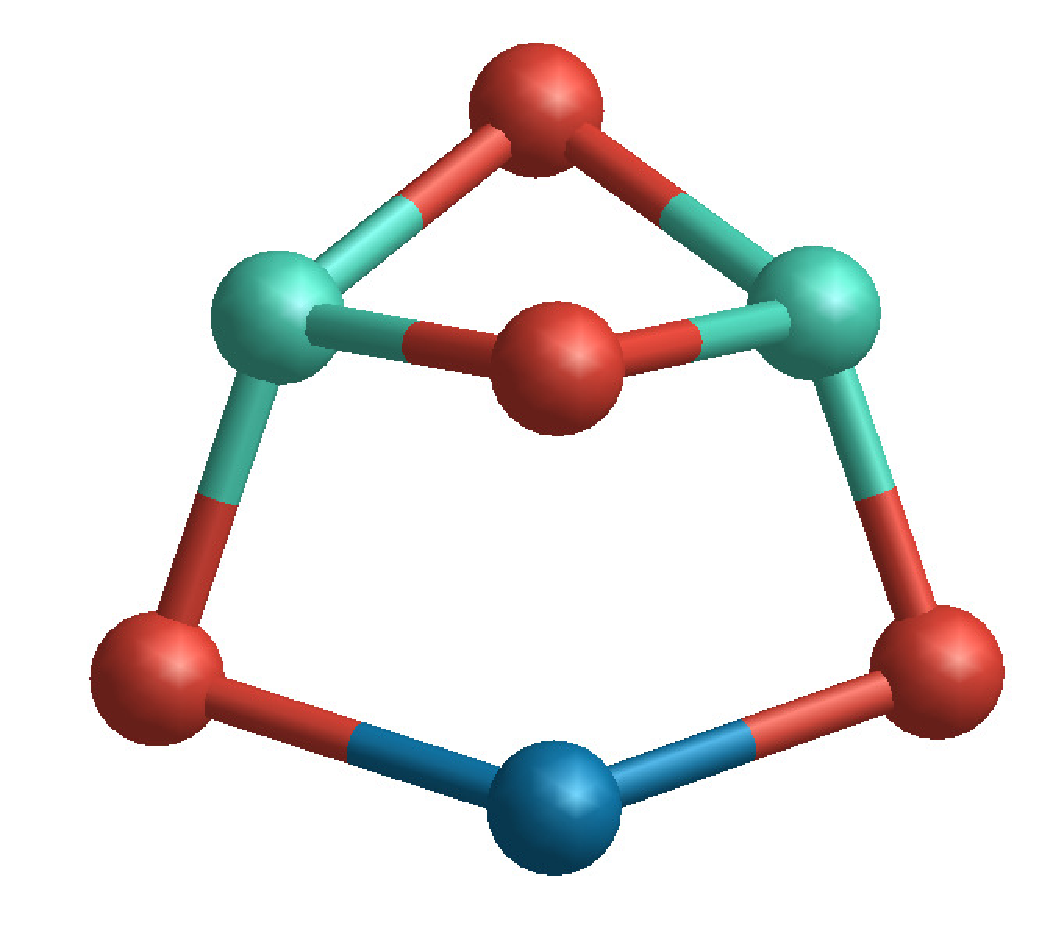} 1A
\includegraphics[width=0.21\textwidth]{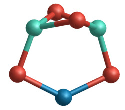} 1B
\caption{Structures of the monomer ($n$=1) GM candidates 1A and 1B for spinel, (MgAl$_2$O$_4$)$_1$, and calcium aluminate (CaAl$_2$O$_4$)$_1$. 
Note that 1B of (CaAl$_2$O$_4$)$_1$ is a transition state.
Mg/Ca atoms are colour-coded in blue, Al atoms in turquoise, and O atoms in red \label{monomer}.}
\end{figure}

The spinel monomer GM candidate (1A) exhibits C$_{s}$ symmetry and is shown in Fig. \ref{monomer}.
The Mg-O bond lengths are 1.980 \AA{} and the Al-O bonds range from 1.703 to 1.811 \AA{}.
The CBS-QB3 binding energy of 1A is 2910 kJ mol$^{-1}$, at the B3LYP/cc-pVTZ level of theory it is lower (2728 kJ mol$^{-1}$). 
We note that isomer 1B, a C$_{2v}$ structure reported by \citep{woodley_ternary}, has a CBS-QB3 relative energy just 10.8 kJ mol$^{-1}$ (B3LYP/cc-pVTZ: 11.0 kJ mol$^{-1}$) above our GM candidate 1A. 
For temperatures above 900 K, structure 1B becomes more favourable according to its Gibbs free energy of formation. 1A and 1B essentially differ by the distance of the Mg cation to the out-of-plane oxygen anion, which is 2.11 \AA{} in 1A and 3.11 \AA{} in 1B, so these structures can be regarded as conformers. 
1B exhibits a very low vibration frequency of 41 cm$^{-1}$, which was identified as a hindered rotation.

Structure 1A also corresponds to the most favourable isomer of the CaAl$_2$O$_4$ monomer with a CBS-QB3 binding energy of 2995 kJ mol$^{-1}$. 
Owing to the mentioned problems of the CBS-QB3 method with Ca (see Sect. \ref{3p1}), we also provide the B3LYP/cc-pVTZ binding energy of 2921 kJ mol$^{-1}$. 
The Ca-O bond lengths of 2.208 \AA{} are larger than the Mg-O bonds in the spinel clusters, as the Ca cation has a larger radius (i.e. an extra shell of electrons). 
The Al-O bonds in 1A range from 1.677 to 1.797 \AA{} and are slightly smaller than for MgAl$_2$O$_4$.
Isomer 1B exhibits an imaginary frequency of 71\textit{i} cm$^{-1}$ for CaAl$_2$O$_4$ monomer and represents a transition state.

\subsubsection{The dimers ($n$=2)}
\begin{figure}[h]
\includegraphics[width=0.24\textwidth]{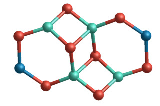}2A
\includegraphics[width=0.19\textwidth]{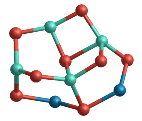}2B
\caption{Dimer ($n$=2) GM candidates 2A and 2B for stoichiometric (MgAl$_2$O$_4$)$_2$ and (CaAl$_2$O$_4$)$_2$ clusters.\label{dimer}}
\end{figure}

The most favourable spinel dimer structure, 2A, is depicted in Fig. \ref{dimer} and has a C$_{i}$ point group symmetry. 
It was previously reported by \citet{woodley_ternary}. 
We find a binding energy of $E_{b}/n$=3114 kJ mol$^{-1}$ 
at the B3LYP/cc-pVTZ level of theory. The Mg-O bond lengths are 1.980 \AA{} and the Al-O bonds range from 1.703 \AA{} to 1.811 \AA{}.
2A shows a large aspect ratio with xyz dimensions of 7.22 \AA{} $\times$ 4.68 \AA{} $\times$ 1.88\AA{}.

A metastable isomer (2B) with a relative energy to 2A of 14 kJ mol$^{-1}$ is found to be the second most favourable spinel dimer structure in our searches. 
For the entire temperature range considered in this study ($T=0-$6000 K) 2B is less favourable than 2A.\
For (CaAl$_2$O$_4$)$_2$ structure 2B is slighly preferred (by 7 kJ/mol) to 2A at $T=$0 K, but 2A becomes favourable for temperatures above room temperature (i.e. 298 K). 
Therefore, we consider 2A to be our best GM candidate for both (CaAl$_2$O$_4$)$_2$ and (MgAl$_2$O$_4$)$_2$. 
The corresponding (CaAl$_2$O$_4$)$_2$ geometry shows a slightly distorted symmetry with respect to the spinel dimer exhibiting two different Ca-O bond lengths of 2.025 \AA{} and 2.082\AA{}, whereas the Al-O bonds range from 1.680 to 1.946 \AA{}.
The calcium aluminate dimer shows a larger B3LYP/cc-pVTZ binding energy ($E_{b}/n$=3271 kJ mol$^{-1}$) than the spinel dimer.\
In addition, different isomers were optimised for mixed calcium-magnesium aluminates, MgCaAl$_4$O$_8$. In this case, 2A corresponds to the GM candidate with a binding energy of $E_{b}/n$=3193 kJ mol$^{-1}$, which lies between the energies for Ca$_2$Al$_4$O$_8$ and Mg$_2$Al$_4$O$_8$, but still closer to the former.

\subsubsection{The trimers ($n$=3)}
\begin{figure}[h]
\includegraphics[width=0.22\textwidth]{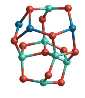}3A
\includegraphics[width=0.22\textwidth]{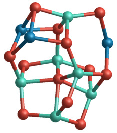}3B
\caption{Trimer ($n$=3) GM candidates for stoichiometric spinel and calcium aluminate clusters, (Mg/CaAl$_2$O$_4$)$_3$.\label{trimer}}
\end{figure}

In Fig. \ref{trimer} the GM candidate of the spinel trimer, (MgAl$_2$O$_{4}$)$_3$, 3A, is shown. We find that structure 3A is lower in energy by 34 kJ mol$^{-1}$ than isomer 3B, which was found by \citet{woodley_ternary}. 
The structures 3A and 3B show an overall similar geometry with differences of an Al-O and a Mg-O bond that are visible at the top of the structures in  Fig. \ref{trimer}. The lowest-energy trimer isomers  
are not symmetric and belong to the C$_1$ space group. 
For (CaAl$_2$O$_{4}$)$_3$ 3B is 11 kJ mol$^{-1}$ more favourable than 3A. For temperatures $>$ 700 K, 3A becomes the preferred geometry. For this reason we report (CaAl$_2$O$_4$)$_3$ results for both 3A and 3B. One of the peculiarities of isomer 3B is that it shows a five-fold coordinated oxygen atom.\
For the mixed Mg/Ca aluminate trimers Mg$_2$CaAl$_6$O$_{12}$ and MgCa$_2$Al$_6$O$_{12}$, 3A corresponds to the favoured isomer. We note that the relative energy between 3A and 3B decreases with increasing Ca content, and becomes negative for (Ca$_3$Al$_6$O$_{12}$)$_3$.  

\subsubsection{The larger clusters ($n$=4$-$7)}
\begin{figure}[h]
\includegraphics[width=0.22\textwidth]{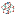}4A
\includegraphics[width=0.22\textwidth]{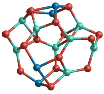}4B
\caption{Tetramer ($n$=4) GM candidates for stoichiometric spinel and calcium aluminate clusters, (Mg/CaAl$_2$O$_4$)$_4$\label{tetramer}.}
\end{figure}

The GM candidates for $n$=4, 4A and 4B, are shown in Fig. \ref{tetramer}. 4A and 4B have very similar structures, but show differences in some bonds and coordinations.
For (MgAl$_2$O$_4$)$_4$, structure 4A is lower in potential energy by 12 kJ mol$^{-1}$ than isomer 4B, which corresponds to the GM candidate found by \citet{woodley_ternary}. For elevated temperatures 4A remains the most favourable cluster isomer. Both isomers, 4A and 4B, show a structural similarity without symmetry.
For (CaAl$_2$O$_4$)$_4$, the optimisation of 4A and 4B leads to a pair of stereoisomers with the same energy.
4A also represents the most favourable geometry for the mixed Ca$_2$Mg$_2$Al$_8$O$_{16}$ cluster species. For temperatures above 5000 K, 4A and 4B have essentially the same free energies. Generally, we note a trend of increasing binding energy of about $\sim$ 40 kJ mol$^{-1}$, when substituting one Mg with one Ca cation.

\begin{figure}[h]
\includegraphics[width=0.22\textwidth]{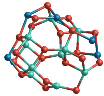}5A
\includegraphics[width=0.22\textwidth]{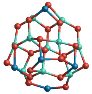}5B
\caption{Pentamer ($n$=5) GM candidates for stoichiometric spinel and calcium aluminate clusters, (Mg/CaAl$_2$O$_4$)$_5$\label{pentamer}.}
\end{figure}

The pentamer ($n$=5) GM candidate 5A is shown in Fig. \ref{pentamer}.
A different C$_s$ symmetric structure, was predicted as GM by \citet{woodley_ternary} for (MgAl$_2$O$_4$)$_5$. This structure was also found in our MC-BH searches. 
However, during the optimisation with the B3LYP/cc-pVTZ method, the symmetry of this isomer was broken, which led to a slightly distorted geometry (5B). 
By imposing symmetry our optimisations did not converge.
The distorted geometry isomer 5B is 32 kJ mol$^{-1}$ above 5A.
For (CaAl$_2$O$_4$)$_5$, 5A lies 73 kJ mol$^{-1}$ below 5B and represents the GM candidate.
Mixed Mg/Ca  aluminates also exhibit 5A as a preferential structure and 
their binding energies scale with the Ca/Mg ratio. The larger the Ca content, the higher is the binding energy $E_{b}/n$ ($\sim$20-30 kJ mol$^{-1}$ per Ca atom).

\begin{figure}[h]
\includegraphics[width=0.22\textwidth]{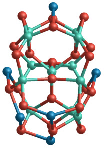}6A
\includegraphics[width=0.22\textwidth]{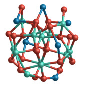}6B
\caption{Hexamer ($n$=6) GM candidates for stoichiometric spinel and calcium aluminate clusters, (Mg/CaAl$_2$O$_4$)$_6$\label{hexamer}.}
\end{figure}

The lowest-energy spinel hexamer ($n$=6) cluster 6A is shown in Fig. \ref{hexamer}. 
6A shows a quasi-symmetric mirror plane, where an Mg atom is replaced by an Al atom on the right hand side of the plane. 
For the isomer reported by \citet{woodley_ternary}, we find a large relative energy of 312 kJ mol$^{-1}$ above 6A. By swapping one Al with one Mg ion in the hexamer structure of \citet{woodley_ternary},
we find the more favourable structure 6B that lies 264 kJ mol$^{-1}$ above 6A. 
For (CaAl$_2$O$_4$)$_6$ 6A corresponds to the most favourable isomer found in our searches and it lies 55 kJ mol$^{-1}$ below 6B. 
The energy difference between the GM candidates of (MgAl$_2$O$_4$)$_6$ and (CaAl$_2$O$_4$)$_6$ is $E_{b}/n$=143 kJ mol$^{-1}$. 
For the mixed Mg/Ca  aluminate hexamers we find a gradual increase in the formation enthalpy with increasing Ca/Mg ratio, consistent with the findings for the other cluster sizes. 
For example, Mg$_3$Ca$_3$Al$_{12}$O$_{24}$ with a Ca:Mg ratio of 1:1 is 68 kJ mol$^{-1}$ higher per unit $n$ than (CaAl$_2$O$_4$)$_6$, but 75 kJ mol$^{-1}$ per unit $n$ lower than (MgAl$_2$O$_4$)$_6$.

\begin{figure}[h]
\includegraphics[width=0.22\textwidth]{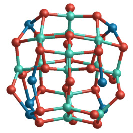}7A
\includegraphics[width=0.22\textwidth]{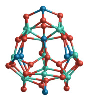}7B
\caption{Heptamer ($n$=7) GM candidates 7A and 7B for stoichiometric spinel and calcium aluminate clusters, (Mg/CaAl$_2$O$_4$)$_7$\label{heptamer}.}
\end{figure}

For the spinel heptamer we find structure 7A to be the lowest-energy isomer (see Fig. \ref{heptamer}).
Similar to 6A, 7A also shows a quasi-mirror plane where an Mg ion is substituted by an Al ion.
Structure 7B reported by \cite{woodley_ternary} shows a potential energy of 66 kJ mol$^{-1}$ above 7A and we find a small imaginary vibration mode with a frequency of 11.32 \textit{i} cm$^{-1}$. The mode could not be attributed to a specific bond stretch or bend, and appears as a collective breathing mode.
For (CaAl$_2$O$_4$)$_7$, 7A has a binding energy $E_{b}/n$ that is 149 kJ mol$^{-1}$ larger than for the (MgAl$_2$O$_4$)$_7$ counterpart. 7B lies 158 kJ mol$^{-1}$ above 7A.
Test calculations of mixed Mg/Ca aluminate clusters confirm the correlation of increased binding energies with increased Ca content for smaller clusters.
The atomic coordinates of the clusters presented in this study can be found in Table \ref{coordinates}.

\subsection{Homogeneous nucleation and the bulk limit}
\label{3p4}
In the following two sub-subsections we summarise the geometric and electrostatic properties of the GM cluster candidates as a function of their size, \textit{n}, and compare the size-dependent trends with the bulk limit that is represented by the crystals spinel and krotite.

\subsubsection{Homogeneous nucleation}
\begin{figure}[h]
\includegraphics[width=0.46\textwidth]{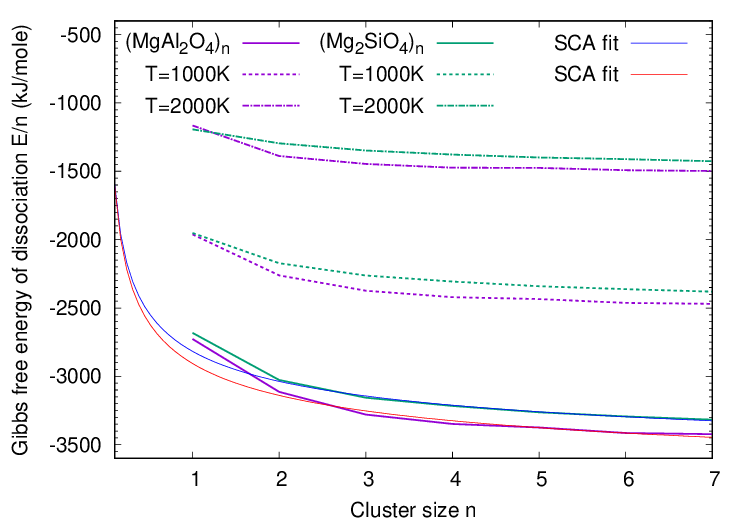}
\caption{Normalised Gibbs free energies of dissociation for the stoichiometric spinel GM clusters, (MgAl$_{2}$O$_{4}$)$_{n}$, as a function of cluster size, \textit{n}, for different temperatures of $T=$0 K (solid line), $T=$1000 K (dashed line), and $T=$2000 K (dash-dotted line). For comparison, the Mg-rich olivine GM candidate clusters, (Mg$_2$SiO$_4$)$_n$, as found in \citet{doi:10.1021/acsearthspacechem.9b00139}, and their free energies are also included.\label{growth}}
\end{figure}

In Fig. \ref{growth}, the Gibbs free energies of dissociation of the GM spinel cluster candidates, normalised to the cluster size, $n$, are shown. 
We illustrate this normalised cluster energy for three different temperatures of $T=$0, 1000, 2000 K. 
At $T=$0 K, the absolute of the Gibbs free energy of dissociation corresponds to the binding energy, and also to the enthalpy of dissociation.
Using the spherical cluster approximation (SCA; see e.g. \cite{Johnston2002}) and excluding the monomer ($n$=1), we fitted the normalised energies in the form of 
\begin{equation}
E(n)/n=a+bn^{-1/3},
\end{equation}
where parameter $a$ corresponds to the normalised bulk energy and parameter $b$ is related to the surface tension. 
For MgAl$_2$O$_4$, we find fitting parameters of 

\begin{equation}
\begin{array}{l}
a=-4037.5\rm{\hspace*{0.1cm}kJ mol^{-1}}, \textit{b}=1131.1 \rm{\hspace*{0.1cm} kJ mol^{-1}\hspace*{0.3cm}for\hspace*{0.1cm}T=0 K} \\
a=-2879.4\rm{\hspace*{0.1cm}kJ mol^{-1}}, \textit{b}=755.7 \rm{\hspace*{0.1cm} kJ mol^{-1}\hspace*{0.5cm}for\hspace*{0.1cm}T=1000 K} \\
a=-1710.0\rm{\hspace*{0.1cm}kJ mol^{-1}}, \textit{b}=393.8 \rm{\hspace*{0.1cm} kJ mol^{-1}\hspace*{0.5cm}for\hspace*{0.1cm} T=2000 K.} 
\end{array}
\end{equation}

The value for $a$ agrees reasonably well with the cohesive energy of 4070 kJ mol$^{-1}$ for crystalline MgAl$_2$O$_4$ at $T=$0 K derived from the JANAF-NIST thermochemical tables\footnote{https://janaf.nist.gov/} \citep{219851}.
When including the monomer in the fitting procedure, we find a larger value for
$a$ of 4241.8 kJ mol$^{-1}$. Moreover, the fittings that include the monomer increasingly overpredict the energies for cluster sizes $n>$5, which reflects the fact that the SCA is derived in the large cluster limit.
With these fitting relations the free energies of larger spinel clusters, whose investigation is computationally very demanding, 
can be predicted.
However, we also note that some cluster sizes (e.g. $n$=3,4) are more favourable than others ($n$=5,7). 
Hence, it is possible that some of the larger clusters with $n>$7 show enhanced stability.

For comparison, we also included Mg-rich olivine clusters, (Mg$_2$SiO$_4$)$_n$, that were studied by \citet{doi:10.1021/acsearthspacechem.9b00139}. 
These silicate clusters can be directly compared to the spinel clusters as they contain the same number of oxygen anions and metal cations per formula unit. 
For coherence and consistency we applied the same functional and basis set (B3LYP/cc-pVTZ) to optimise the silicate clusters.
Except for the monomer ($n=$1) and large temperatures ($T>$1300 K), the spinel clusters are more favourable than their corresponding silicate clusters for all shown cluster sizes ($n$=1$-$7) and temperatures ($T=$0, 1000, 2000 K).  
The SCA fitting parameters for Mg$_2$SiO$_4$ silicate clusters are 
\begin{equation}
\begin{array}{l}
a=-3882.4\rm{\hspace*{0.1cm}kJ mol^{-1}}, \textit{b}=1065.1\rm{\hspace*{0.1cm}kJ mol^{-1}\hspace*{0.3cm} for\hspace*{0.1cm} T=0 K} \\ 
a=-2783.2\rm{\hspace*{0.1cm}kJ mol^{-1}}, \textit{b}=761.1 \rm{\hspace*{0.1cm}kJ mol^{-1}\hspace*{0.5cm} for\hspace*{0.1cm} T=1000 K} \\
a=-1676.9\rm{\hspace*{0.1cm}kJ mol^{-1}}, \textit{b}=477.5 \rm{\hspace*{0.1cm}kJ mol^{-1}\hspace*{0.5cm} for\hspace*{0.1cm} T=2000 K},
\end{array}
\end{equation}
where $a$ at $T=$0K is in very good agreement with the value of 3888 kJ mol$^{-1}$ derived from JANAF-NIST for crystalline magnesium-rich olivine. As for spinel clusters, the inclusion of the monomer leads to a more negative value of $a$ (-4036.55 kJ mol$^{-1}$) and to a worse fit for larger cluster sizes.

\begin{figure}[h]
\includegraphics[width=0.46\textwidth]{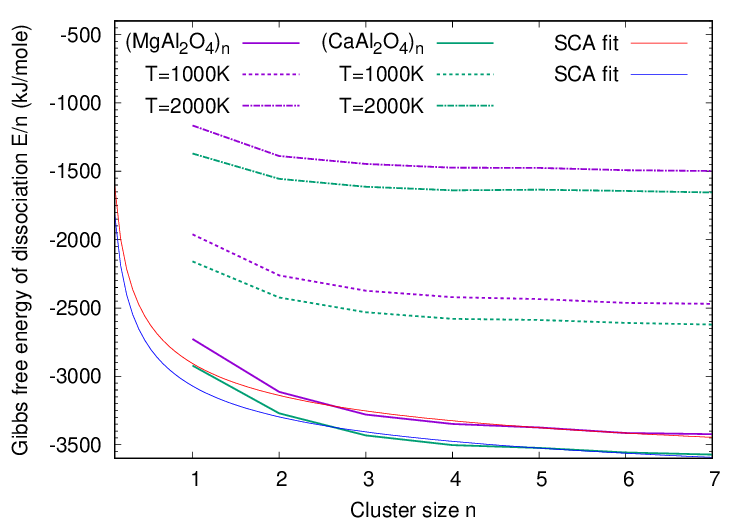}
\caption{Normalised Gibbs free energies of dissociation for the stoichiometric krotite clusters, (CaAl$_{2}$O$_{4}$)$_{n}$, and spinel clusters, (MgAl$_{2}$O$_{4}$)$_{n}$, as a function of cluster size, $n,$ for different temperatures of $T=$0 K (solid line), $T=$1000 K (dashed line), and $T=$2000 K (dash-dotted line).\label{casp}} 
\end{figure}

In addition to the silicates, the MgAl$_2$O$_4$ cluster energies are compared to those of Ca aluminates (i.e. CaAl$_2$O$_4$). 
In Fig. \ref{casp}, the Gibbs free energies of dissociation of the aluminate clusters with respect to their atomic components, are compared for different temperatures as a function of the cluster size, \textit{n}. 
The (CaAl$_2$O$_4$)$_n$ clusters are more favourable than their corresponding Mg spinels.
This can partially be explained by the stronger Ca-O bond (414 kJ mol$^{-1}$) in comparison with the Mg-O bond (260 kJ mol$^{-1}$). 

The SCA fit for (CaAl$_2$O$_4$)$_n$, $n$=2-7, results in fitting parameters of 
\begin{equation}
\begin{array}{l}
a=-4164.6\rm{\hspace*{0.1cm}kJ mol^{-1}}, \textit{b}=1093.4 \rm{kJ mol^{-1} \hspace*{0.3cm}for\hspace*{0.1cm} T=0 K}\\ 
a=-3005.6\rm{\hspace*{0.1cm}kJ mol^{-1}}, \textit{b}=710.2 \rm{kJ mol^{-1} \hspace*{0.5cm}for\hspace*{0.1cm} T=1000 K}\\ 
a=-1838.3\rm{\hspace*{0.1cm}kJ mol^{-1}}, \textit{b}=341.5 \rm{kJ mol^{-1} \hspace*{0.5cm}for\hspace*{0.1cm} T=2000 K}.
\end{array}
\end{equation}

To our knowledge, there is no thermo-chemical data for crystalline krotite in the literature. Different high pressure polymorphs of CaAl$_2$O$_4$ show cohesive energies of 6.11$-$6.15 eV/atom corresponding to 4127$-$4154 kJ mol$^{-1}$ \citep{QI2020120259}, which agrees fairly well with our fitted constant $a$.
Comparing these fits to those for the spinel clusters and extrapolating to larger sizes, we find that clusters of (MgAl$_2$O$_4$)$_n$ are less favourable than those of (CaAl$_2$O$_4$)$_n$ for all sizes $n$.
Since this extrapolation is not based on actual cluster data, it should be handled with care.
It should be noted here that clusters with mixed Mg-Ca stoichiometries also exist, for example MgCaAl$_4$O$_8$. 
As shown in Sect. \ref{3p3} the energies of these mixed clusters lie in between the values for (MgAl$_2$O$_4$)$_n$ and (CaAl$_2$O$_4$)$_n$. 
The thermochemical tables of the GM candidate clusters derived in this study can be found in Table 
\ref{thermotable}.

\subsubsection{The bulk limit}
\begin{figure}[h]
\includegraphics[width=0.46\textwidth]{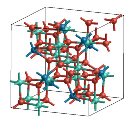}
\caption{Unit cell of crystalline spinel, MgAl$_2$O$_4$, with the cell parameters a=b=c=8.089\AA{} and $\alpha=\beta=\gamma=90^{\circ}$  adopted from \citet{Finger1986} \label{unitcell}}
\end{figure}

Crystalline spinel shows a T$^2_d$ symmetry in the cubic crystal system with unit cell parameters a=b=c=8.089 \AA{} and $\alpha$=$\beta$=$\gamma$=90$^{\circ}$ \citep{Finger1986}. The spinel unit cell is illustrated in Fig. \ref{unitcell}.
In this form, Al atoms are 6-coordinated, O atoms 4-coordinated, and Mg 4-coordinated, respectively. The bond distances are Mg-O 1.889 \AA{} and Al-O 2.058 \AA{}.

Generally, the AlO and MgO bond lengths as well as the atomic coordinations increase as a function of cluster size, as can be seen in Table \ref{geom}. 
This is not unexpected as the fraction of `surface' atoms decreases with cluster size and, therefore, the average coordination and the bond lengths increases. However, the increase is not strictly monotonic, for example $\overline{d}$(AlO) for $n$=4 and $\overline{d}$(MgO) for $n$=2 represent outliers. Also, the average coordinations of the $n$=7 GM cluster are lower than those for $n$=6. 
Moreover, individual bond length and atomic coordinations in the clusters can differ from both their average values and the bulk value (i.e. $n\rightarrow \infty$). For example, structure 3B bears a 5-fold coordinated oxygen, which is larger than the coordination in the bulk.\\

\begin{figure}[h]
\includegraphics[width=0.46\textwidth]{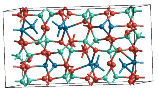}
\caption{Unit cell of crystalline krotite, CaAl$_2$O$_4$. The cell parameters --- a=8.700\AA{}, b=8.099\AA{}, c=15.217\AA,{} and $\alpha$=$\gamma$=90$^{\circ}$, $\beta$=90.188$^{\circ}$ --- are adapted from \citet{10.2138am.2011.3693} \label{krokite_unitcell}.}
\end{figure}

In Fig. \ref{krokite_unitcell} the krotite unit cell is shown. It exhibits a $\Gamma _{m}C_{2h}^{5}$ symmetry in the monoclinic crystal system with unit cell parameters a=8.700\AA{}, b=8.099\AA{}, c=15.217\AA{} and $\alpha$=$\gamma$=90$^{\circ}$, $\beta$=90.188$^{\circ}$  \citep{10.2138am.2011.3693}

\begin{table}[h]
\caption{Geometric properties and binding energies of the (MgAl$_2$O$_4$)$_n$ GM candidate clusters.\label{geom}}
\begin{tabular}{c  r r   r r r r} 
$n$ & $\overline{d}$(AlO)  \tablefootnote[1]{mean interatomic distances in \AA{}}        &  $\overline{d}$(MgO)           & $\overline{c}$(Al) \tablefootnote[2]{mean coordination numbers} & $\overline{c}$(Mg) & $\overline{c}$(O)     & E$_b/n$ \tablefootnote[3]{normalised binding energies (in kJ mol$^{-1}$)}\\
\hline
1 &  1.7533 &  1.9557  & 3.00  &   2.00  &   2.00  & 2728 \\
2 &  1.7811 &  1.8285  & 3.50  &   2.00  &   2.25  & 3114 \\
3 &  1.7908 &  1.9794  & 3.67  &   3.00  &   2.58  & 3280 \\
4 &  1.7872 &  1.9810  & 3.75  &   3.00  &   2.63  & 3328 \\
5 &  1.8016 &  1.9899  & 3.90  &   3.00  &   2.70  & 3375 \\
6 &  1.8145 &  1.9986  & 4.08  &   3.17  &   2.83  & 3415 \\
7 &  1.8345 &  1.9827  & 4.29  &   2.57  &   2.79  & 3423 \\
\hline
$\infty$ & 1.9120 &  1.9160  & 6.00  &   4.00  &   4.00 & 4070 \tablefootnote[4]{derived from JANF-NIST\citep{219851}} \\
\end{tabular}
\end{table}

In Table \ref{geomCa} the geometric properties of the (CaAl$_2$O$_4$)$_n$ GM candidate clusters
are shown. The average Al-O bond lengths are very similar to those of the MgAl$_2$O$_4$ clusters and increase with cluster size, except for $n$=4. 
For Ca-O, we chose a cutoff for the maximum bond length of 2.35\AA{}, and for Al-O and Mg-O a maximum of 2.00\AA{}.
While the coordination of the Al cations and the O anions increase almost monotonically with cluster size, the Ca coordination does not follow this trend.

\begin{table}[h]
\caption{Geometric properties and binding energies of the (CaAl$_2$O$_4$)$_n$ GM candidate clusters.\label{geomCa}}
\begin{tabular}{c   r r   r r r r} 
$n$ & $\overline{d}$(AlO)  \tablefootnote[1]{mean interatomic distances in \AA{}} &  $\overline{d}$(CaO)           & $\overline{c}$(Al) \tablefootnote[2]{mean coordination numbers} & $\overline{c}$(Ca) & $\overline{c}$(O) & E$_b/n$\tablefootnote[3]{normalised binding energies (in kJ mol$^{-1}$)} \\
\hline
1 & 1.748  & 2.207  & 3.00  & 3.00    & 2.25   & 2922 \\
2 & 1.778  & 2.053  & 3.50  & 2.00    & 2.25    & 3271 \\
3(3A) & 1.790 & 2.218  & 3.50  & 3.00    & 2.50   & 3432 \\
3(3B) & 1.814 & 2.183  & 4.00  & 2.33    & 2.58  & 3435 \\
4 &  1.783 & 2.232  & 3.75  & 3.00    & 2.63  & 3503 \\
5 & 1.809  & 2.206  & 4.00  & 2.80    & 2.70  & 3524 \\
6 & 1.813  & 2.237  & 4.08  & 2.67    & 2.71  & 3558 \\
7 & 1.839  & 2.233  & 4.43  & 2.43    & 2.82  & 3572 \\
\hline
$\infty$ & 1.754  & 2.342  & 4.00  & 4.00   & 3.00 & 4165 \tablefootnote[5]{derived from SCA fit (this study)} \\
\end{tabular}
\end{table}

Electrostatic properties of the spinel clusters and the bulk limit ($n=\infty$) are given in Table \ref{elecstat}.
For the mean Al charge, an increasing trend with size \textit{n} can be seen. The average charges of the Mg and O ions do not, however,  follow a clear trend with respect to the cluster size. 
Generally, all Al and Mg ions are positively charged cations, and the O ions are negatively charged anions.
Apart for the symmetric dimer ($n$=2) the spinel clusters exhibit considerable dipole moments with a maximum value of 7.32 Debye for $n$=3. Therefore, we predict that our GM spinel cluster candidates should be detectable by IR spectroscopy, if they are present.
The highest molecular orbital--lowest unoccupied molecular orbital (HOMO-LUMO) gap of the spinel GM clusters ranges from 3.88 to 5.11 eV and is just below the band gap of the crystalline bulk spinel of $\sim$5.11 eV \citep{Pilania2020}. 
The ionisation potentials range between 7.36 and 9.24 eV and generally decrease with cluster size. 

\begin{table}[h]
\caption{Electrostatic properties of the (MgAl$_2$O$_4$)$_n$ GM candidate clusters.\label{elecstat}}  
\begin{tabular}{c   r r r   r r   r r}
$n$ & $\overline{q}$(Al) \tablefootnote[1]{average Mulliken charge (in e)} & $\overline{q}$(Mg) & $\overline{q}$(O) & Dipole \tablefootnote[2]{total moment (in Debye)} &  Gap \tablefootnote[3]{HOMO-LUMO gap (in eV)} & E$_i^v$ \tablefootnote[4]{vertical and adiabatic ionisation energies (in eV)} & E$_i^a$   \\  
\hline                                              
1 &  0.583  & 0.838 & -0.501  &    2.94 &  3.88 & 9.24 & 8.94    \\ 
2 &  0.608  & 0.631 & -0.462  &    0.00 &  4.68 & 9.00 & 8.74    \\ 
3 &  0.590  & 0.871 & -0.513  &    7.33 &  4.80 & 8.97 & 8.40    \\ 
4 &  0.606  & 0.826 & -0.509  &    1.93 &  4.57 & 8.64 & 8.40    \\ 
5 &  0.623  & 0.839 & -0.521  &    5.79 &  4.62 & 8.57 & 7.87    \\ 
6 &  0.702  & 0.851 & -0.564  &    2.98 &  5.11 & 8.62 & 8.01    \\ 
7 &  0.725  & 0.799 & -0.562  &    3.42 &  4.88 & 8.44 & 7.36    \\
\hline
$\infty$ &    &      &    &  & 5.11 \tablefootnote[5]{experimental value from \citet{Pilania2020}}  & & \\           
\end{tabular}
\end{table}

In Table \ref{elecstatCa} the electrostatic properties of the (CaAl$_2$O$_4$)$_n$ GM candidate clusters are displayed. The Al cation charges generally increase with cluster sizes n for both cluster families. For CaAl$_2$O$_4$ the Al charges are slightly larger than for MgAl$_2$O$_4$, except for $n$=1,2.
Ca charges are considerably more positive than the Mg charges of spinel clusters. The oxygen anions show slightly more negative average charges for krotite than for spinel clusters. 
The larger atomically partitioned charges in krotite clusters is also reflected in their larger dipole moments. For $n\ge$3 the (CaAl$_2$O$_4$)$_n$ clusters exhibit dipole moments with large values $>$8 Debye making them suitable targets for IR observations.
The HOMO-LUMO gaps of the two cluster families are comparable, though the range in (CaAl$_2$O$_4$)$_n$ is narrower (i.e. 4.19-4.84 eV). In this range also the band gap of crystalline calcium aluminate ($\sim$4.54 eV) is located \citep{doi:10.1021/acs.chemmater.5b00288}.
The vertical and adiabatic ionisation energies are slightly lower for (CaAl$_2$O$_4$)$_n$ and generally decrease with cluster size, \textit{n}.

\begin{table}[h]
\caption{Electrostatic properties of the (CaAl$_2$O$_4$)$_n$ GM candidate clusters. They include (1) cluster size, $n$, (2) average Mulliken charge (in e), (5) the total dipole moment (in Debye), and (6) the HOMO-LUMO gap (in eV), (7) the vertical and (8) the adiabatic ionisation energies (in eV).
\label{elecstatCa}}
\begin{tabular}{c  r r r  r r  r r}
$n$ & $\overline{q}$(Al) \tablefootnote[1]{average Mulliken charge (in e)} &  $\overline{q}$(Ca) & $\overline{q}$(O)     &    Dipole \tablefootnote[2]{total moment (in Debye)} &  Gap \tablefootnote[3]{HOMO-LUMO gap (in eV)}& E$_i^v$ \tablefootnote[4]{vertical and adiabatic ionisation energies (in eV)} & E$_i^a$   \\  
\hline                                              
1 & 0.565   & 1.053 & -0.546  & 4.90    & 4.19  & 8.78 & 8.30 \\ 
2 & 0.586   & 0.982 & -0.538  & 0.00    & 4.69  & 8.24 & 7.93 \\ 
3A & 0.657  & 1.018 & -0.583  & 10.99   & 4.84  & 8.37 & 7.85  \\ 
3B & 0.698  & 0.967 & -0.591  & 8.28    & 4.30  & 7.99 & 7.19  \\
4 & 0.666   & 0.999 & -0.583  & 7.13    & 4.64  & 7.95 & 7.50   \\ 
5 & 0.701   & 0.983 & -0.596  & 8.82    & 4.46  & 7.81 & 7.26 \\ 
6 & 0.844   & 0.876 & -0.641  & 8.93    & 4.51  & 7.62 & 7.17 \\ 
7 & 0.831   & 0.821 & -0.6205 & 8.31    & 4.58  & 7.50 & 6.44 \\
\hline
$\infty$ &     &    &    &  & 4.54 \tablefootnote[6]{experimental value from \citet{doi:10.1021/acs.chemmater.5b00288}}  & & \\           
\end{tabular}
\end{table}


\subsection{Harmonic spectra}
\label{3p5}
\begin{figure}[h]
\includegraphics[width=0.46\textwidth]{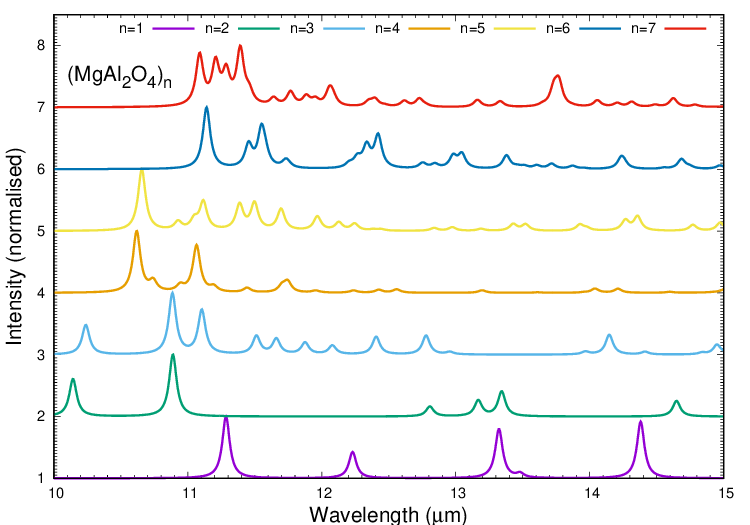}
\caption{Harmonic vibrational spectra of (MgAl$_{2}$O$_{4}$)$_n$ clusters as a function of wavelength. The normalised IR intensities are plotted with a Lorentzian profile and a full width at half maximum of 0.033 $\mu$m \label{spectraMg}.}
\end{figure}

Clusters exhibit 3N-6 vibrational modes, where N is the number of constituent atoms.
Therefore, the spinel monomer ($n$=1) shows 15 modes and the heptamer ($n$=7) 141 modes. 
We note that for the monomer ($n$=1) two of 15 and for the dimer ($n$=2) 18 of 36 vibrational modes are IR inactive; all larger considered spinel GM cluster candidates have only IR active modes.
The most intense vibrational emission lines of the spinel family are found in a wavelength range between 10.5 and 11.5 $\mu$m (see Fig. \ref{spectraMg}). 
With regard to the 13 $\mu$m feature, emissions from the monomer ($n$=1), the dimer ($n$=2) and the hexamer ($n$=6) are predicted. However, their relative intensities are rather low.

\begin{figure}[h]
\includegraphics[width=0.46\textwidth]{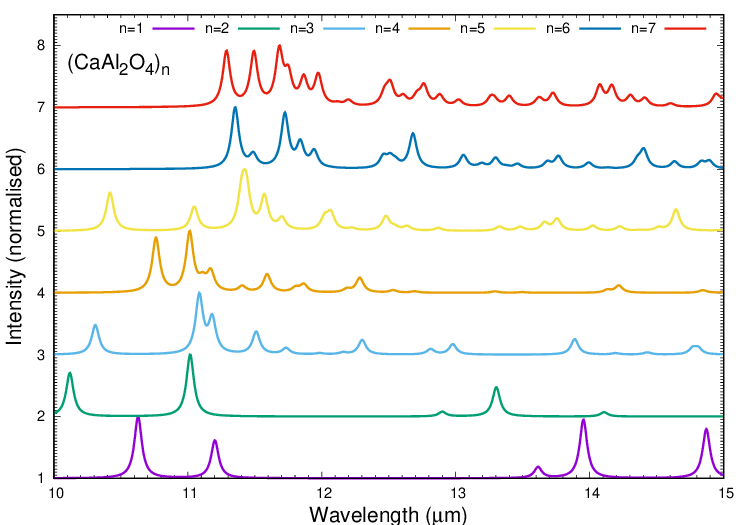}
\caption{Harmonic vibrational spectra of (CaAl$_{2}$O$_{4}$)$_n$ clusters as a function of wavelength. The normalised IR intensities are plotted with a Lorentzian profile and a full width at half maximum of 0.033 $\mu$m \label{caspectraH}.}
\end{figure}

The harmonic vibrational spectra of the (CaAl$_{2}$O$_{4}$)$_n$ GM cluster candidates are shown in Fig. \ref{caspectraH}. 
The most intense vibrational emissions occur between 10.5 and 12.0 $\mu$m.
There are common spectral features with (MgAl$_{2}$O$_{4}$)$_n$ clusters (see e.g. for $n$=2), but also differences are apparent (see e.g. $n$=5).
Some peaks of (CaAl$_{2}$O$_{4}$)$_n$ are slightly shifted towards longer wavelengths.

We note that several assumptions and approximations are made for the vibrational IR spectra presented in this study. 
They include the harmonic approximation that neglects anharmonic and temperature effects, but also the simplification that only GM candidates of (Mg/CaAl$_2$O$_4$)$_n$, $n$=1$-$7 stoichiometry contribute to the IR emission. 
However, a detailed treatment of the IR spectra in stellar atmospheres with radiative transfer modelling is beyond the scope of this paper.

\section{Discussion}
\label{4}
The present paper, along with Paper I, demonstrates that the kinetic formation of alumina clusters and particles is efficient. A kinetic CaAl$_2$O$_4$ synthesis is viable if pulsational shocks are excluded, while MgAl$_2$O$_4$ does not form in O-rich circumstellar envelopes.

The analysis of pre-solar oxygen-rich silicate grains in meteorites revealed that many grains contain a certain fraction of Ca and Al (see e.g. \citet{2008ApJ...682.1450N, 2010ApJ...719..166N, 2010ApJ...714.1624B}). Moreover, \citet{2018GeCoA.221..255L} found an unusual large AGB stardust grain with an inner core that contains Al and Ca, but no Si. Therefore, alumina and Ca- and Mg-bearing aluminates do indeed appear to serve, at least in some cases, as seed nuclei for larger silicate grains, which is in agreement with our studies.

\begin{figure}
\includegraphics[width=0.49\textwidth]{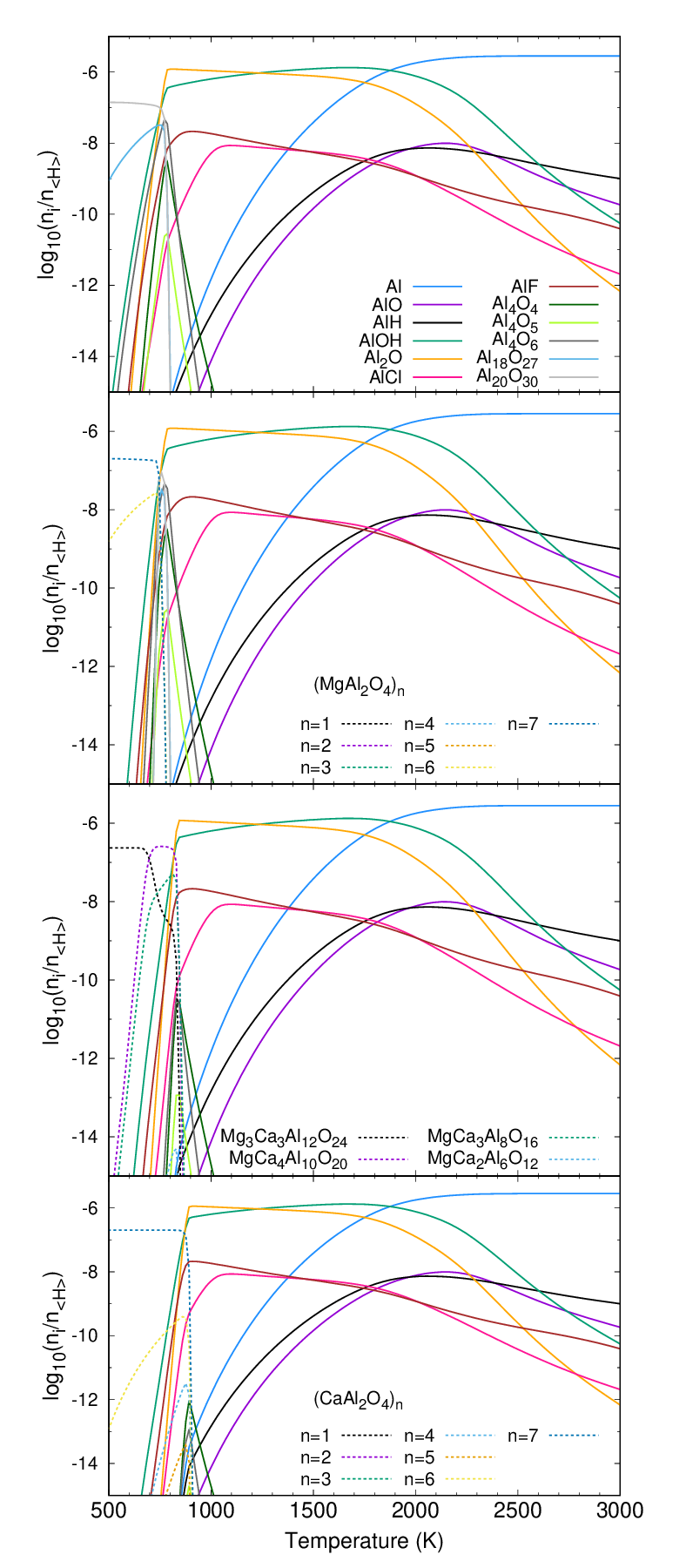}
\caption{Chemical equilibrium calculations performed at p=110 dyne cm$^{-2}$ and with the software GGchem \citep{2018A&A...614A...1W} for a large gas-phase mixture with 865 species and including clusters of  (Al$_2$O$_3$)$_n$, $n$=1$-$10 (top); (MgAl$_2$O$_4$)$_n$, $n$=1$-$7 and (Al$_2$O$_3$)$_n$, $n$=1$-$10 ({second row);} (Mg$_x$Ca$_{1-x}$Al$_2$O$_4$)$_n$, x$\in$ (0..1], (MgAl$_2$O$_4$)$_n$, $n$=1$-$7, and Al$_2$O$_3$)$_n$, $n$=1$-$10 (third row); and (CaAl$_2$O$_4$)$_n$, (Mg$_x$Ca$_{1-x}$Al$_2$O$_4$)$_n$, x$\in$(0,1], (MgAl$_2$O$_4$)$_n$, $n$=1$-$7, and (Al$_2$O$_3$)$_n$, $n$=1$-$10\label{chemeq}. (bottom).}
\end{figure}

Chemical equilibrium calculations show a different picture.
To showcase these differences, we applied conditions very similar  to those used in 
the present kinetic study: the same initial elemental composition, a pressure corresponding to the photospheric gas density in  the MIRA models, and a temperature range of 500 K and 3000 K  
in the chemical equilibrium code GGchem \citep{2018A&A...614A...1W}.
We find that none of the considered cluster families (i.e. alumina, spinel, mixed Mg/Ca aluminates, and krotite) sustains above a temperature of $T=$900 K (see Fig. \ref{chemeq}) 
The top panel includes a gas-phase mixture of 865 species and the thermochemical data of (Al$_2$O$_3$)$_n$, $n$=1$-$10 clusters, whereas the second panel additionally includes the (MgAl$_2$O$_4$)$_n$, $n$=1$-$7 clusters reported in this study (see Table \ref{dgfits} for the corresponding fitting coefficients).
The largest alumina cluster, (Al$_2$O$_3$)$_{10}$, predominates only for temperatures below 850 K. 
The inclusion of different (MgAl$_2$O$_4$)$_n$ cluster sizes leads to a situation, where the largest cluster with $n=$7 essentially replaces (Al$_2$O$_3$)$_{10}$.
This is in clear contrast to the results of our chemical-kinetic study predicting alumina clusters as primary seed particles and no spinel formation in any of the considered models.
If additionally mixed Mg/Ca aluminate clusters and (CaAl$_2$O$_4$)$_n$ are included (see the bottom panel of Fig. \ref{chemeq}), we find that the Ca-rich aluminates dominate the Al equilibrium chemistry, which is in agreement with the thermochemical energies derived in this study.

Generally, chemical equilibrium 
abundances are valid for conditions that remain constant for an infinite time. However, this 
is not the case in many highly dynamical astrophysical environments, where active dust formation takes place.
Therefore, a chemical-kinetic approach accounting for reaction timescales and barriers represents a more correct approach than using chemical equilibrium. 
In equilibrium, the species concentrations do not change by definition and their formation routes cannot be traced. More important, in an equilibrium approach the role of reaction barriers and unstable intermediates is ignored.

It is also worth noting that non-thermal effects can have a significant impact on the chemistry in circumstellar envelopes. 
The importance of vibrational (or internal) thermal non-equilibrium for circumstellar dust formation was postulated in the past decades \citep{1981ApJ...247..925N,1998A&A...337..847P}.
As has been shown in \citet{D2FD00025C}, clusters with large dipole moments can efficiently lower their internal temperatures via spontaneous and stimulated photon emission. 
In turn, the lower internal temperatures lead to dissociation rates that are significantly reduced, and aid the nucleation to proceed at an accelerated pace.
Moreover, small temperature differences between different sizes of the same cluster species can affect the corresponding nucleation rates considerably \citep{2023A&A...671A169K}. In this study we do not account for different vibrational and translational temperatures of the species, or cluster sizes. 
This means that in the pulsating models the effect of rapidly changing temperatures is implemented 
simultaneously for all considered species.

Regarding comparison with recent observations, we note that the model results for the prevalent molecules CO, H$_2$O, and OH agree well \citep{Maerker,2023A&A...674A.125B}. 
For the Al chemistry, the modelled abundances are in accordance with observations with the exception 
of AlOH, where model abundances exceed the observed values by one to two orders of magnitude. This fact was previously noted and discussed in detail in Paper I.

So far, no Mg- or Ca-bearing molecules have been detected around oxygen-rich AGB stars.
Owing to the relatively large dipole moments of their oxides (MgO and CaO), sulfides (MgS and CaS), and hydroxides (MgOH and CaOH), it is unlikely that these molecules are very abundant. 
As has been noted by \citet{Agundez_2020}, neutral atoms are likely the main reservoir of magnesium and calcium in AGB atmospheres.
We largely agree with this conclusion, but do not exclude the possibility of Ca/Mg being in the form of Ca$^+$ and Mg$^+$ cations or part of nascent dust grains.

The kinetic network presented in this study includes termolecular and bimolecular neutral-neutral reactions. Atomic and molecular ions, however, are not considered in this study. 
The first ionisation energy of Mg atoms is 7.65 eV, which is similar to those of stoichiometric (MgO)$_n$ clusters ranging from 7.1 to 8.2 eV \citep{2021Univ....7..243G}. The ionisation potentials of the spinel and krotite clusters presented in this study are in a similar range of $\sim$ 7$-$10 eV (see Tables \ref{elecstat} and \ref{elecstatCa}).
Atomic Al has the lowest relevant ionisation potential of 5.99 eV.
These energies are fairly large compared to the thermal energies at the photosphere of AGB stars of typcially 0.15$-$0.3 eV. 
Dredged-up $^{26}$Al with a half-life of $\sim$ 717 000 years might represent a more efficient source of ionisation than temperature, and is expected to be become particularly important in C-rich AGB stars that have experienced several dredge-up episodes. In addition, some AGB stars show significant UV emission, which is possibly caused by chromospheric activity and could lead to a partial ionisation of circumstellar matter \citep{2017ApJ...841...33M}. Nevertheless, a relatively low ionisation fraction is expected in these environments. However, even a low ionisation degree can impact the chemistry since ion-molecule rates are typically orders of magnitude faster than neutral-neutral reactions.

It is also possible that the spinel and krotite nucleation does not proceed via the monomer as presumed in this study, but via different stoichiometries or different kick-starter species (i.e. heterogeneously). 
The inclusion of Mg in clusters represents a major challenge in modelling bottom-up dust nucleation in oxygen-rich environments. This is not only the case for spinel, but also for Mg-rich silicates of olivine and pyroxene stoichiometry, as well as for Mg-bearing titanates that are affected by this problem \citep{doi:10.1098/rsta.2012.0335}.
In contrast, Ca can be incorporated more easily in clusters under certain circumstances as was shown in this study. 
These circumstances include the absence of pulsational shocks and the exclusion of the competing alumina nucleation.

Comparing our kinetic results to classical nucleation descriptions used in, for example, \citet{2022A&A...668A..35S}, we find formal `monomeric' radii of 2.164 \AA{} for Al$_2$O$_3$, 2.505 \AA{} for MgAl$_2$O$_4$, and 2.771 \AA{} for CaAl$_2$O$_4$, respectively. 
At a temperature of $T=$1000 K, surface tensions of 2.027$\times$10$^{-4}$ J cm$^{-2}$ for Al$_2$O$_3$ and 1.741$\times$10$^{-4}$ J cm$^{-2}$ for MgAl$_2$O$_4$ can be derived
using the cluster energies derived in this study; for CaAl$_2$O$_4$ no value for the surface tension is provided, since thermodynamic information on the crystalline bulk is lacking.

From the Gibbs free energy extrapolation of the SCA fit, we find that krotite clusters are more favourable than their Mg-rich counterparts for all sizes, $n$. 
We note that there are several caveats in the interpretation of this result. 
Foremost, this result is a fitted extrapolation for $n>$7 and does not rely on actually calculated or measured cluster data. Also, this approximation does not account for particularly favourable `magic' cluster sizes, or energetically unfavourable nucleation bottlenecks. 
Second, as shown for small cluster sizes $n$=1$-$7, it is likely that also mixed Ca/Mg aluminate clusters exist with energies that are in between pure krotite and spinel clusters.
Although Ca-rich clusters are favoured for all cluster sizes, Mg is about an order of magnitude more abundant than Ca and could be incorporated in the clusters at some stage. For these reasons, we believe that the larger Ca ions can successively be replaced by smaller and more abundant Mg ions.

As clusters are typically intermediate in size between gas-phase molecules and solid bulk material, their respective vibrational spectra show properties that are in between discrete molecular line emissions and broad dust features. This is a consequence of their number of degrees of freedom that scale with the cluster size (i.e. the number of atoms that they contain). 
As the clusters presented here have several Al$-$O and Mg/Ca$-$O bonds of different lengths, their respective intense stretching modes cover a range of wavelengths, leading to a non-discrete spectrum. Yet the cluster spectra are not as broadened over a large wavelength range as the emissions of sub-micron-sized dust grains. 
As for alumina clusters, the most intense vibrational modes are located in the 10.5$-$11.5 $\mu$m wavelength range. These intense modes are attributed to Al$-$O stretchings, whereas the Mg/Ca-O modes are more modest and occur at slightly longer wavelengths (i.e. 10.5$-$12.0 $\mu$m). 
Therefore, these clusters are unlikely to be carriers of the 13 $\mu$m feature, which is expected to arise from larger particles.
The emissions of larger cluster particles become largely independent of the interior composition and will gradually evolve towards a black body radiation.
Generally, the IR spectrum of circumstellar dust shells is dominated by fully grown grains that could cover the spectral signatures of nucleation species with smaller sizes.
In reality, the cluster spectra are influenced by anharmonicities and elevated temperatures that can lead not only to broadening, but also to shifts in wavelengths, appearance and/or disappearance of certain peaks and asymmetries in the intensities \citep{doi:10.1021/acsearthspacechem.0c00341}.

\section{Summary}
\label{5}
In this study we explored several kinetic pathways for the formation of spinel (MgAl$_2$O$_4$) and krotite (CaAl$_2$O$_4$) monomers.
Under certain conditions, including the absence of pulsational shocks and the exclusion of alumina cluster nucleation,
the krotite monomer, CaAl$_2$O$_4$, and its hydroxylated form, CaAl$_2$O$_3$(OH$_2$), can be produced in significant amounts, up to a fractional abundance of 2$\times10^{-8}$. 
This corresponds to slightly less than 1\% of the global budget of the elements aluminium and calcium, and would result in a krotite dust-to-gas mass ratio of 1.5 $\times10^{-6}$, which is two orders of magnitude lower than the alumina dust-to-gas mass ratio of 1.1 $\times10^{-4}$.
However, the kinetic formation of the MgAl$_2$O$_4$ monomer represents a major challenge under circumstellar conditions. 
In particular, the inclusion of Mg in spinel and silicates is inefficient since the reverse reactions proceed at a higher speed. In contrast, the nucleation of alumina is efficient close to the star (1 R$_{\star}$$-$2.5 R$_{\star}$) and is only barely affected by the inclusion of the magnesium and calcium aluminate chemistry. Therefore, alumina remains the most likely seed particle candidate according to our physico-chemical models.

Presuming the existence of monomers including (MgAl$_2$O$_4$)$_1$, the subsequent cluster growth is energetically favourable for temperatures in the dust condensation zone (i.e. T$<$2000 K). Extensive global optimisation searches were performed to derive the energies and structures of the most favourable cluster isomers for (Mg/CaAl$_2$O$_4$)$_n$, $n$=1$-$7, including mixed Mg/Ca aluminates. For cluster sizes $n$=3$-$7, hitherto unreported GM candidates were revealed. Some of these lowest-energy isomers show large dipole moments and are therefore potentially suitable for future IR observations.
From the thermodynamic properties of the (sub-)nanometre-sized clusters presented in this study, we predict a stability sequence in which CaAl$_2$O$_4$ clusters are the most favourable species, followed by mixed Ca/Mg aluminate clusters, MgAl$_2$O$_4$, and olivinic silicates. 
The harmonic vibrational spectra of the clusters cannot account for the commonly observed 13 $\mu$m feature in circumstellar envelopes, which likely arises due to grown (sub-)micrometre-sized alumina and aluminate dust grains.

Instead, the most intense vibrational modes are found in a wavelength regime between 10.5 and 11.5 $\mu$m for (MgAl$_2$O$_4$)$_n$ and between $10.5$ and 12 $\mu$m for (CaAl$_2$O$_4$)$_n$, $n=1-7$ clusters.\\

\section*{Acknowledgements}
Financial support from the Knut and Alice Wallenberg foundation is gratefully acknowledged through grant nr. KAW 2020.0081. 
We acknowledge the CINECA award under the ISCRA initiative, for the availability of high performance computing resources and support.
The computations involved the Swedish National Infrastructure for Computing (SNIC) at Chalmers Centre for Computational Science and Engineering (C3SE) partially funded by the Swedish Research Council through grant no. 2018-05973. The computations were partially enabled by resources provided by the National Academic Infrastructure for Supercomputing in Sweden (NAISS) partially funded by the Swedish Research Council through grant agreement no. 2022-06725.
JMCP acknowledges support from the UK Science Technology and Facilities Council (grant ST/T000287/1), and computing time on the Leeds ARC4 supercomputer.
STB acknowledges support from the MICINN funded project grants: PID2021-
127957NB-I00, TED2021-132550B-C21 and CEX2021-001202-M and grant 2021SGR00354 funded by the Generalitat de Catalunya.
LD acknowledges support from  the KU Leuven C1 grant MAESTRO C16/17/007 and from  the Research Foundation Flanders (FWO) grant G099720N.

\bibliographystyle{aa}
\bibliography{biblio}

\begin{thebibliography}{70}
\expandafter\ifx\csname natexlab\endcsname\relax\def\natexlab#1{#1}\fi

\bibitem[{{Ag\'undez} {et~al.}(2020){Ag\'undez}, I., {de Andres, P. L.},
  {Cernicharo, J.}, \& {Mart\'{\i}n-Gago, J. A.}}]{Agundez_2020}
{Ag\'undez}, M., I., M., {de Andres, P. L.}, {Cernicharo, J.}, \&
  {Mart\'{\i}n-Gago, J. A.} 2020, A\&A, 637, A59

\bibitem[{Andersson {et~al.}(2023)Andersson, Gobrecht, \&
  Valero}]{10.3389/fspas.2023.1135156}
Andersson, S., Gobrecht, D., \& Valero, R. 2023, Frontiers in Astronomy and
  Space Sciences, 10

\bibitem[{{Baudry} {et~al.}(2023){Baudry}, {Wong}, {Etoka}, {Richards},
  {M{\"u}ller}, {Herpin}, {Danilovich}, {Gray}, {Wallstr{\"o}m}, {Gobrecht},
  {Khouri}, {Decin}, {Gottlieb}, {Menten}, {Homan}, {Millar}, {Montarg{\`e}s},
  {Pimpanuwat}, {Plane}, \& {Kervella}}]{2023A&A...674A.125B}
{Baudry}, A., {Wong}, K.~T., {Etoka}, S., {et~al.} 2023, \aap, 674, A125

\bibitem[{{Becke}(1993)}]{1993JChPh..98.1372B}
{Becke}, A.~D. 1993, jcp, 98, 1372

\bibitem[{{Bose} {et~al.}(2010){Bose}, {Floss}, \&
  {Stadermann}}]{2010ApJ...714.1624B}
{Bose}, M., {Floss}, C., \& {Stadermann}, F.~J. 2010, \apj, 714, 1624

\bibitem[{{Boulangier} {et~al.}(2019){Boulangier}, {Gobrecht}, {Decin}, {de
  Koter}, \& {Yates}}]{2019MNRAS.tmp.2040B}
{Boulangier}, J., {Gobrecht}, D., {Decin}, L., {de Koter}, A., \& {Yates}, J.
  2019, mnras, 2040

\bibitem[{Bromley \& Flikkema(2005)}]{PhysRevLett.95.185505}
Bromley, S.~T. \& Flikkema, E. 2005, Phys. Rev. Lett., 95, 185505

\bibitem[{Bromley {et~al.}(2016)Bromley, Gomez~Martin, \& Plane}]{C6CP03629E}
Bromley, S.~T., Gomez~Martin, J.~C., \& Plane, J. M.~C. 2016, Phys. Chem. Chem.
  Phys., 18, 26913

\bibitem[{Chase(1998)}]{219851}
Chase, M. 1998, NIST-JANAF Thermochemical Tables, 4th Edition (American
  Institute of Physics, -1)

\bibitem[{Connelly {et~al.}(2012)Connelly, Bizzarro, Krot, Åke Nordlund,
  Wielandt, \& Ivanova}]{doi:10.1126/science.1226919}
Connelly, J.~N., Bizzarro, M., Krot, A.~N., {et~al.} 2012, Science, 338, 651

\bibitem[{{Cristallo} {et~al.}(2015){Cristallo}, {Straniero}, {Piersanti}, \&
  {Gobrecht}}]{2015ApJS..219...40C}
{Cristallo}, S., {Straniero}, O., {Piersanti}, L., \& {Gobrecht}, D. 2015,
  \apjs, 219, 40

\bibitem[{Decin {et~al.}(2018)Decin, Danilovich, Gobrecht, Plane, Richards,
  Gottlieb, \& Lee}]{Decin_2018}
Decin, L., Danilovich, T., Gobrecht, D., {et~al.} 2018, The Astrophysical
  Journal, 855, 113

\bibitem[{{Decin} {et~al.}(2010){Decin}, {De Beck}, {Br\"unken, S.}, {M\"uller,
  H. S. P.}, {Menten, K. M.}, {Kim, H.}, {Willacy, K.}, {de Koter, A.}, \&
  {Wyrowski, F.}}]{Decin2010}
{Decin}, L., {De Beck}, E., {Br\"unken, S.}, {et~al.} 2010, A\&A, 516, A69

\bibitem[{{Decin} {et~al.}(2017){Decin}, {Richards, A. M. S.}, {Waters, L. B.
  F. M.}, {Danilovich, T.}, {Gobrecht, D.}, {Khouri, T.}, {Homan, W.}, {Bakker,
  J. M.}, {Van de Sande, M.}, {Nuth, J. A.}, \& {De Beck, E.}}]{refId0D}
{Decin}, L., {Richards, A. M. S.}, {Waters, L. B. F. M.}, {et~al.} 2017, A\&A,
  608, A55

\bibitem[{Douglas {et~al.}(2022)Douglas, Gobrecht, \&
  Plane}]{10.1093/mnras/stac1684}
Douglas, K.~M., Gobrecht, D., \& Plane, J. M.~C. 2022, Monthly Notices of the
  Royal Astronomical Society, 515, 99

\bibitem[{Escatllar {et~al.}(2019)Escatllar, Lazaukas, Woodley, \&
  Bromley}]{doi:10.1021/acsearthspacechem.9b00139}
Escatllar, A.~M., Lazaukas, T., Woodley, S.~M., \& Bromley, S.~T. 2019, ACS
  Earth and Space Chemistry, 3, 2390

\bibitem[{{Fabian} {et~al.}(2001){Fabian}, {Posch}, {Mutschke}, {Kerschbaum},
  \& {Dorschner}}]{2001A&A...373.1125F}
{Fabian}, D., {Posch}, T., {Mutschke}, H., {Kerschbaum}, F., \& {Dorschner}, J.
  2001, \aap, 373, 1125

\bibitem[{Finger {et~al.}(1986)Finger, Hazen, \& Hofmeister}]{Finger1986}
Finger, L.~W., Hazen, R.~M., \& Hofmeister, A.~M. 1986, Physics and Chemistry
  of Minerals, 13, 215

\bibitem[{{Freytag} {et~al.}(2017){Freytag}, {Liljegren}, \&
  {H{\"o}fner}}]{2017A&A...600A.137F}
{Freytag}, B., {Liljegren}, S., \& {H{\"o}fner}, S. 2017, \aap, 600, A137

\bibitem[{Frisch {et~al.}(2016)Frisch, Trucks, Schlegel, Scuseria, Robb,
  Cheeseman, Scalmani, Barone, Petersson, Nakatsuji, Li, Caricato, Marenich,
  Bloino, Janesko, Gomperts, Mennucci, Hratchian, Ortiz, Izmaylov, Sonnenberg,
  Williams-Young, Ding, Lipparini, Egidi, Goings, Peng, Petrone, Henderson,
  Ranasinghe, Zakrzewski, Gao, Rega, Zheng, Liang, Hada, Ehara, Toyota, Fukuda,
  Hasegawa, Ishida, Nakajima, Honda, Kitao, Nakai, Vreven, Throssell,
  Montgomery, Peralta, Ogliaro, Bearpark, Heyd, Brothers, Kudin, Staroverov,
  Keith, Kobayashi, Normand, Raghavachari, Rendell, Burant, Iyengar, Tomasi,
  Cossi, Millam, Klene, Adamo, Cammi, Ochterski, Martin, Morokuma, Farkas,
  Foresman, \& Fox}]{g16}
Frisch, M.~J., Trucks, G.~W., Schlegel, H.~B., {et~al.} 2016, Gaussian˜16
  {R}evision {C}.01, gaussian Inc. Wallingford CT

\bibitem[{{Gail} {et~al.}(2013){Gail}, {Wetzel}, {Pucci}, \&
  {Tamanai}}]{2013A&A...555A.119G}
{Gail}, H.~P., {Wetzel}, S., {Pucci}, A., \& {Tamanai}, A. 2013, aap, 555, A119

\bibitem[{Georgievskii \& Klippenstein(2005)}]{doi:10.1063/1.1899603}
Georgievskii, Y. \& Klippenstein, S.~J. 2005, The Journal of Chemical Physics,
  122, 194103

\bibitem[{Gilbert \& Smith(1990)}]{Gilbert1990TheoryOU}
Gilbert, R. \& Smith, S. 1990, in Theory of Unimolecular and Recombination
  Reactions

\bibitem[{Glowacki {et~al.}(2012)Glowacki, Liang, Morley, Pilling, \&
  Robertson}]{doi:10.1021/jp3051033}
Glowacki, D.~R., Liang, C.-H., Morley, C., Pilling, M.~J., \& Robertson, S.~H.
  2012, The Journal of Physical Chemistry A, 116, 9545, pMID: 22905697

\bibitem[{{Gobrecht} {et~al.}(2016){Gobrecht}, {Cherchneff}, {Sarangi},
  {Plane}, \& {Bromley}}]{2016A&A...585A...6G}
{Gobrecht}, D., {Cherchneff}, I., {Sarangi}, A., {Plane}, J.~M.~C., \&
  {Bromley}, S.~T. 2016, aap, 585, A6

\bibitem[{{Gobrecht} {et~al.}(2022){Gobrecht}, {Plane}, {Bromley, Stefan T.},
  {Decin, Leen}, {Cristallo, Sergio}, \& {Sekaran, Sanjay}}]{Gobrecht_alumina}
{Gobrecht}, D., {Plane}, {Bromley, Stefan T.}, {et~al.} 2022, A\&A, 658, A167

\bibitem[{{Gobrecht} {et~al.}(2021){Gobrecht}, {Sindel}, {Lecoq-Molinos}, \&
  {Decin}}]{2021Univ....7..243G}
{Gobrecht}, D., {Sindel}, J.~P., {Lecoq-Molinos}, H., \& {Decin}, L. 2021,
  Universe, 7, 243

\bibitem[{{Gottlieb} {et~al.}(2022){Gottlieb}, {Decin}, {Richards}, {De
  Ceuster}, {Homan}, {Wallstr{\"o}m}, {Danilovich}, {Millar}, {Montarg{\`e}s},
  {Wong}, {McDonald}, {Baudry}, {Bolte}, {Cannon}, {De Beck}, {de Koter}, {El
  Mellah}, {Etoka}, {Gobrecht}, {Gray}, {Herpin}, {Jeste}, {Kervella},
  {Khouri}, {Lagadec}, {Maes}, {Malfait}, {Menten}, {M{\"u}ller}, {Pimpanuwat},
  {Plane}, {Sahai}, {Van de Sande}, {Waters}, {Yates}, \&
  {Zijlstra}}]{2022A&A...660A..94G}
{Gottlieb}, C.~A., {Decin}, L., {Richards}, A.~M.~S., {et~al.} 2022, \aap, 660,
  A94

\bibitem[{Goumans \& Bromley(2012)}]{doi:10.1111/j.1365-2966.2011.20255.x}
Goumans, T. P.~M. \& Bromley, S.~T. 2012, Monthly Notices of the Royal
  Astronomical Society, 420, 3344

\bibitem[{Goumans \& Bromley(2013)}]{doi:10.1098/rsta.2011.0580}
Goumans, T. P.~M. \& Bromley, S.~T. 2013, Philosophical Transactions of the
  Royal Society A: Mathematical, Physical and Engineering Sciences, 371,
  20110580

\bibitem[{Guiu {et~al.}(2021)Guiu, Escatllar, \&
  Bromley}]{doi:10.1021/acsearthspacechem.0c00341}
Guiu, J.~M., Escatllar, A.~M., \& Bromley, S.~T. 2021, ACS Earth and Space
  Chemistry, 5, 812

\bibitem[{{Gyngard} {et~al.}(2010){Gyngard}, {Zinner}, {Nittler}, {Morgand},
  {Stadermann}, \& {Mairin Hynes}}]{2010ApJ...717..107G}
{Gyngard}, F., {Zinner}, E., {Nittler}, L.~R., {et~al.} 2010, apj, 717, 107

\bibitem[{{Henning}(2010)}]{2010LNP...815.....H}
{Henning}, T., ed. 2010, Lecture Notes in Physics, Berlin Springer Verlag, Vol.
  815, {Astromineralogy}

\bibitem[{{Hindmarsh}(2019)}]{2019ascl.soft05021H}
{Hindmarsh}, A.~C. 2019, {ODEPACK: Ordinary differential equation solver
  library}, Astrophysics Source Code Library, record ascl:1905.021

\bibitem[{{H{\"o}fner} \& {Olofsson}(2018)}]{2018A&ARv..26....1H}
{H{\"o}fner}, S. \& {Olofsson}, H. 2018, aapr, 26, 1

\bibitem[{{Johnston}(2002)}]{Johnston2002}
{Johnston}, R.~L. 2002, {Atomic and Molecular Clusters } (CRC Press)

\bibitem[{{Kiefer} {et~al.}(2023){Kiefer}, {Gobrecht}, {Decin}, \&
  {Helling}}]{2023A&A...671A169K}
{Kiefer}, S., {Gobrecht}, D., {Decin}, L., \& {Helling}, C. 2023, \aap, 671,
  A169

\bibitem[{Klippenstein {et~al.}(1988)Klippenstein, Khundkar, Zewail, \&
  Marcus}]{klippenstein1988application}
Klippenstein, S., Khundkar, L., Zewail, A., \& Marcus, R. 1988, The Journal of
  chemical physics, 89, 4761

\bibitem[{{Koehler} {et~al.}(1997){Koehler}, {Gail}, \&
  {Sedlmayr}}]{1997A&A...320..553K}
{Koehler}, T.~M., {Gail}, H.~P., \& {Sedlmayr}, E. 1997, \aap, 320, 553

\bibitem[{{Leitner} {et~al.}(2018){Leitner}, {Hoppe}, {Floss}, {Hillion}, \&
  {Henkel}}]{2018GeCoA.221..255L}
{Leitner}, J., {Hoppe}, P., {Floss}, C., {Hillion}, F., \& {Henkel}, T. 2018,
  \gca, 221, 255

\bibitem[{{Ma} {et~al.}(2011){Ma}, {Kampf}, {Connolly}, Beckett, Rossman,
  Smith, \& Schrader}]{10.2138am.2011.3693}
{Ma}, C., {Kampf}, A.~R., {Connolly}, {Harold}~C., J., {et~al.} 2011, American
  Mineralogist, 96, 709

\bibitem[{{Maercker} {et~al.}(2016){Maercker}, {Danilovich}, {Olofsson}, {De
  Beck}, {Justtanont}, {Lombaert}, \& {Royer}}]{Maerker}
{Maercker}, M., {Danilovich}, T., {Olofsson}, H., {et~al.} 2016, A\&A, 591, A44

\bibitem[{{Mangan} {et~al.}(2021){Mangan}, {Douglas}, {Gobrecht}, {Lade},
  {Decin}, \& {Plane}}]{Mangan2021}
{Mangan}, T., {Douglas}, K., {Gobrecht}, D., {et~al.} 2021, {ACS Earth and
  Space Chemistry} submitted, ref: sp-2021-00225y

\bibitem[{{Montarg{\`e}s} {et~al.}(2023){Montarg{\`e}s}, {Cannon}, {de Koter},
  {Khouri}, {Lagadec}, {Kervella}, {Decin}, {McDonald}, {Homan}, {Waters},
  {Sahai}, {Gottlieb}, {Malfait}, {Maes}, {Pimpanuwat}, {Jeste}, {Danilovich},
  {De Ceuster}, {Van de Sande}, {Gobrecht}, {Wallstr{\"o}m}, {Wong}, {El
  Mellah}, {Bolte}, {Herpin}, {Richards}, {Baudry}, {Etoka}, {Gray}, {Millar},
  {Menten}, {M{\"u}ller}, {Plane}, {Yates}, \&
  {Zijlstra}}]{2023A&A...671A..96M}
{Montarg{\`e}s}, M., {Cannon}, E., {de Koter}, A., {et~al.} 2023, \aap, 671,
  A96

\bibitem[{{Montez} {et~al.}(2017){Montez}, {Ramstedt}, {Kastner}, {Vlemmings},
  \& {Sanchez}}]{2017ApJ...841...33M}
{Montez}, Rodolfo, J., {Ramstedt}, S., {Kastner}, J.~H., {Vlemmings}, W., \&
  {Sanchez}, E. 2017, \apj, 841, 33

\bibitem[{Montgomery {et~al.}(2000)Montgomery, Frisch, Ochterski, \&
  Petersson}]{doi:10.1063/1.481224}
Montgomery, J.~A., Frisch, M.~J., Ochterski, J.~W., \& Petersson, G.~A. 2000,
  The Journal of Chemical Physics, 112, 6532

\bibitem[{Mostefaoui \& Hoppe(2004)}]{Mostefaoui_2004}
Mostefaoui, S. \& Hoppe, P. 2004, The Astrophysical Journal, 613, L149

\bibitem[{{Nguyen} {et~al.}(2010){Nguyen}, {Nittler}, {Stadermann}, {Stroud},
  \& {Alexander}}]{2010ApJ...719..166N}
{Nguyen}, A.~N., {Nittler}, L.~R., {Stadermann}, F.~J., {Stroud}, R.~M., \&
  {Alexander}, C. M.~O. 2010, \apj, 719, 166

\bibitem[{{Nittler} {et~al.}(2008){Nittler}, {Alexander}, {Gallino}, {Hoppe},
  {Nguyen}, {Stadermann}, \& {Zinner}}]{2008ApJ...682.1450N}
{Nittler}, L.~R., {Alexander}, C.~M.~O., {Gallino}, R., {et~al.} 2008, apj,
  682, 1450

\bibitem[{{Nuth} \& {Donn}(1981)}]{1981ApJ...247..925N}
{Nuth}, J.~A. \& {Donn}, B. 1981, \apj, 247, 925

\bibitem[{{Patzer} {et~al.}(1998){Patzer}, {Gauger}, \&
  {Sedlmayr}}]{1998A&A...337..847P}
{Patzer}, A.~B.~C., {Gauger}, A., \& {Sedlmayr}, E. 1998, \aap, 337, 847

\bibitem[{Pilania {et~al.}(2020)Pilania, Kocevski, Valdez, Kreller, \&
  Uberuaga}]{Pilania2020}
Pilania, G., Kocevski, V., Valdez, J.~A., Kreller, C.~R., \& Uberuaga, B.~P.
  2020, Communications Materials, 1, 84

\bibitem[{Plane(2013)}]{doi:10.1098/rsta.2012.0335}
Plane, J. M.~C. 2013, Philosophical Transactions of the Royal Society A:
  Mathematical, Physical and Engineering Sciences, 371, 20120335

\bibitem[{Plane \& Robertson(2022)}]{D2FD00025C}
Plane, J. M.~C. \& Robertson, S.~H. 2022, Faraday Discuss., 238, 461

\bibitem[{Pople {et~al.}(1987)Pople, Head‐Gordon, \&
  Raghavachari}]{10.1063/1.453520}
Pople, J.~A., Head‐Gordon, M., \& Raghavachari, K. 1987, The Journal of
  Chemical Physics, 87, 5968

\bibitem[{{Posch} {et~al.}(1999){Posch}, {Kerschbaum}, {Mutschke}, {Fabian},
  {Dorschner}, \& {Hron}}]{1999A&A...352..609P}
{Posch}, T., {Kerschbaum}, F., {Mutschke}, H., {et~al.} 1999, aap, 352, 609

\bibitem[{Purvis \& Bartlett(1982)}]{10.1063/1.443164}
Purvis, George~D., I. \& Bartlett, R.~J. 1982, The Journal of Chemical Physics,
  76, 1910

\bibitem[{Qi {et~al.}(2020)Qi, Spagnoli, \& Fourie}]{QI2020120259}
Qi, C., Spagnoli, D., \& Fourie, A. 2020, Construction and Building Materials,
  264, 120259

\bibitem[{Qu {et~al.}(2015)Qu, Zhang, Wang, Zhou, \&
  Zeng}]{doi:10.1021/acs.chemmater.5b00288}
Qu, B., Zhang, B., Wang, L., Zhou, R., \& Zeng, X.~C. 2015, Chemistry of
  Materials, 27, 2195

\bibitem[{Rietmeijer {et~al.}(1999)Rietmeijer, III, \&
  Karner}]{Rietmeijer_1999}
Rietmeijer, F. J.~M., III, J. A.~N., \& Karner, J.~M. 1999, The Astrophysical
  Journal, 527, 395

\bibitem[{Schäfer {et~al.}(1992)Schäfer, Horn, \& Ahlrichs}]{cc-pVTZ}
Schäfer, A., Horn, H., \& Ahlrichs, R. 1992, The Journal of Chemical Physics,
  97, 2571

\bibitem[{{Sindel} {et~al.}(2022){Sindel}, {Gobrecht}, {Helling}, \&
  {Decin}}]{2022A&A...668A..35S}
{Sindel}, J.~P., {Gobrecht}, D., {Helling}, C., \& {Decin}, L. 2022, \aap, 668,
  A35

\bibitem[{{Sloan} {et~al.}(2003){Sloan}, {Kraemer}, {Goebel}, \&
  {Price}}]{2003ApJ...594..483S}
{Sloan}, G.~C., {Kraemer}, K.~E., {Goebel}, J.~H., \& {Price}, S.~D. 2003, apj,
  594, 483

\bibitem[{{Stock} {et~al.}(2018){Stock}, {Kitzmann}, {Patzer}, \&
  {Sedlmayr}}]{2018MNRAS.479..865S}
{Stock}, J.~W., {Kitzmann}, D., {Patzer}, A. B.~C., \& {Sedlmayr}, E. 2018,
  \mnras, 479, 865

\bibitem[{{Tielens}(2022)}]{2022FrASS...9.8217T}
{Tielens}, A.~G.~G.~M. 2022, Frontiers in Astronomy and Space Sciences, 9,
  908217

\bibitem[{Varshni(1957)}]{varshni1957comparative}
Varshni, Y.~P. 1957, Reviews of Modern Physics, 29, 664

\bibitem[{{Wales} \& {Doye}(1998)}]{1998cond.mat..3344W}
{Wales}, D. \& {Doye}, J. 1998, arXiv:cond-mat/9803344

\bibitem[{{Woitke} {et~al.}(2018){Woitke}, {Helling}, {Hunter}, {Millard},
  {Turner}, {Worters}, {Blecic}, \& {Stock}}]{2018A&A...614A...1W}
{Woitke}, P., {Helling}, C., {Hunter}, G.~H., {et~al.} 2018, \aap, 614, A1

\bibitem[{Woodley(2009)}]{woodley_ternary}
Woodley, S. 2009, Materials and Manufacturing Processes, 24, 255

\bibitem[{{Zega} {et~al.}(2014){Zega}, {Nittler}, {Gyngard}, {Alexander},
  {Stroud}, \& {Zinner}}]{2014GeCoA.124..152Z}
{Zega}, T.~J., {Nittler}, L.~R., {Gyngard}, F., {et~al.} 2014, \gca, 124, 152

\end{thebibliography}

\begin{appendix}
\section{Molecular systems of AlOMg, AlOCa, AlOMgH, and AlOCaH}
\label{A1}
The PES of the small molecular systems AlOMg, AlOMgH, AlOCa, and AlOCaH are explored at the B3LYP/cc-pVTZ level of theory to find the stationary points (i.e. reactants, intermediates, saddle points, and products). 
For consistency the harmonic vibrational analysis is conducted at the same level of theory. 
To obtain more reliable energy barrier heights, single point calculations at the CCSD(T)/cc-pVTZ level of quantum theory are performed \citep{10.1063/1.443164,10.1063/1.453520}. We note that all the 
energies reported here are corrected for B3LYP zero-point energies. 
The reactions involved in the present study are chemically activated processes, in which two reactants form an adduct. The newly formed adduct can either dissociate back to the reactants or can form new products. 
On the other hand, the stabilisation of the adduct can compete with the reactive channels. 
Reactions of this type are generally pressure dependent. 
We used RRKM theory along with a Master Equation analysis to treat such systems.

\begin{figure}[h]
\includegraphics[width=0.48\textwidth]{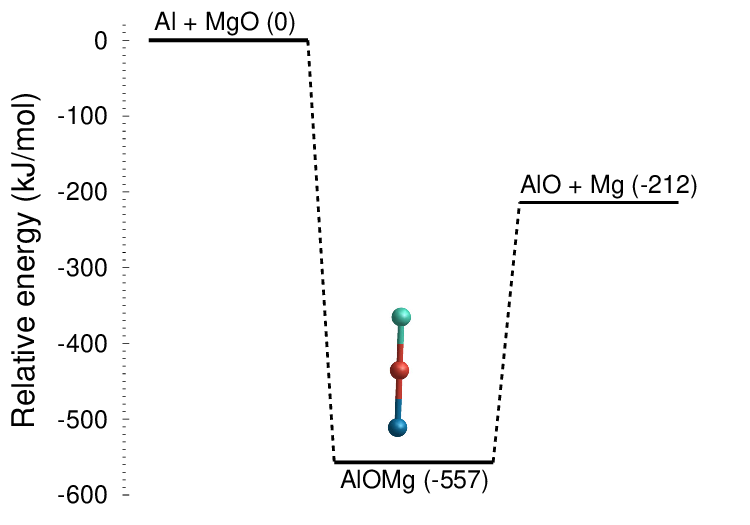}
\includegraphics[width=0.48\textwidth]{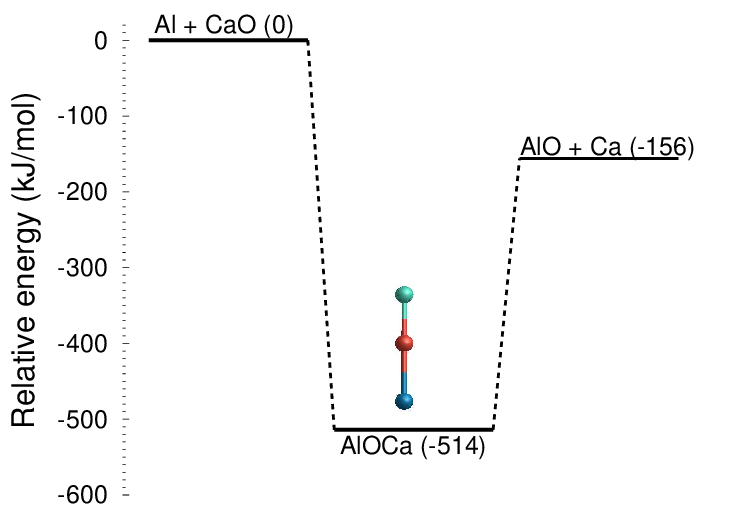}
\includegraphics[width=0.48\textwidth]{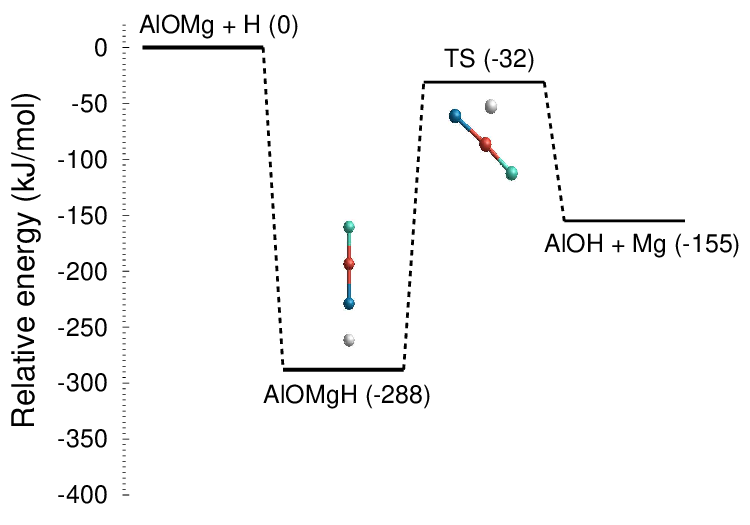}
\includegraphics[width=0.48\textwidth]{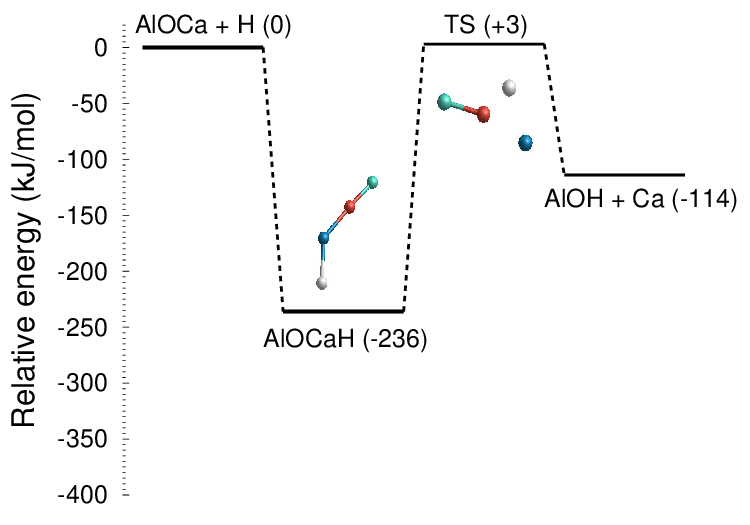}
\caption{Potential energy diagrams for the molecular systems of AlOMg (top panel), AlOCa (second panel), AlOMgH (third panel), and AlOCaH (bottom panel)\label{pes_rasoul}.}
\end{figure}

Some reaction channels show no intrinsic barrier height. 
In these cases variable reaction coordinate-transition state theory (VRC-TST), in which the transition states are considered loose and their positions along the reaction coordinates are located variationally, is applied to obtain the corresponding rate constants. 
In the present study, the bond distance of interest in the adduct is 
varied in a stepwise manner
towards
the corresponding 
reagents and products
at CCSD(T)/cc-pVTZ level. 
Then, the potential energies at these points are fitted to the Varshni equation, which is used in the VRC-TST calculations \citep{varshni1957comparative}. The Variflex code was used for this purpose \citep{klippenstein1988application}.\
The PES of the Al-O-Mg/Ca system 
is shown in Fig. \ref{pes_rasoul}.

This system includes reaction \ref{alo+m},
\begin{equation}
\rm{AlO + Mg/Ca \rightleftharpoons MgO/CaO + Al} 
\end{equation}with $\Delta_{r}$H(0K)= $+$212 kJ mol$^{-1}$ and $\Delta_{r}$H(0K)= $+$156 kJ mol$^{-1}$ 
for the formation of MgO and CaO, respectively. 
These endothermic reaction enthalpies are the result of the stronger Al-O bond as compared with the alkaline earth metal oxides MgO and CaO. We note that the PESs shown in Fig. \ref{pes_rasoul} are explored at the B3LYP/cc-pVTZ level of theory, and CCSD(T)/cc-pVTZ single point energies are then calculated to obtain more reliable potential energies for the stationary points. 
The Al-O-Mg/Ca molecular system can also result in a termolecular association reaction \ref{term},

\begin{equation}
\rm{AlO + Mg/Ca + M \rightleftharpoons AlOMg/AlOCa + M} 
\end{equation}
with $\Delta_{r}$H(0K)= $-$345 kJ mol$^{-1}$ for the formation of AlOMg, and  $\Delta_{r}$H(0K)= $-$358 kJ mol$^{-1}$ for AlOCa.
Although the AlOMg and AlOCa formations are very exothermic, they proceed as termolecular reactions at circumstellar densities and are consequently slow.\\

Owing to the presence of circumstellar AlOH we also studied the Al-O-H-Mg/Ca system (see the bottom panel of Fig. \ref{pes_rasoul}):

\begin{equation}
\rm{AlOH + Mg/Ca  \rightleftharpoons AlOMg/AlOCa + H}   
.\end{equation}

The reaction of AlOH with atomic Mg is endothermic by 155 kJ mol$^{-1}$ with respect to the formation of AlOMg+H. 
The formation of the AlOMgH adduct is exothermic by 133 kJ mol$^{-1}$ with respect to AlOH+Mg, but involves a tight transition transition state that lies 123 kJ mol$^{-1}$ above the reactants.
The situation is similar for the AlOCaH system. AlOCaH can form exothermically (-122 kJ mol$^{-1}$) from AlOH+Ca, but involves a tight transition state 117 kJ mol$^{-1}$ higher than the reagents. 

\begin{table*}[!ht]
\caption{Reaction rate network. (1) the reaction number, (2) the reaction, (3) the CBS-QB3 heat of reaction (enthalpy) at $T=$ 0 K, (4) the reaction rate with the pre-exponential rate constant A given as a(-b) = a $\times$ 10$^{-b}$ (for bimolecular reactions in units of cm$^3$ s$^{-1}$, for termolecular
reactions in units of cm$^6$ s$^{-1}$), and (5) the reference or method of calculation. Reactions 1-166 are adopted from \citet{Gobrecht_alumina} and reaction 167-184 from \citet{Decin_2018} \label{chemnetwork}.}
\begin{tabular}{r l r l l}
no. & Reaction & $\Delta$ H$_r$(0K) & Rate k & Reference  \\
\hline
\rule{0pt}{4ex}
 185 & {Al + MgO $\rightarrow$ AlO + Mg}  & -241  &3.96(-10)(T/300)$^{-0.14}\exp(6.5/T)$ & RRKM \\ 
 186 & {AlO + Mg $\rightarrow$ Al +  MgO} & 241  &1.07(-09)(T/300)$^{-0.05}\exp(-25732/T)$ & RRKM \\
 187 & {AlOH + Mg $\rightarrow$  AlOMg + H} & 157   &5.32(-10)(T/300)$^{0.20}\exp(-18539.8/T)$ & RRKM \\
 188 & {AlOMg + H  $\rightarrow$  AlOH  +  Mg} & -157  &2.50(-10)(T/300)$^{-0.28}\exp(5.8/T)$ & RRKM \\
 189 & {AlO + Mg+ H$_2$ $\rightarrow$ AlOMg +  H$_2$} & -326 &5.32(-30)(T/300)$^{-3.42}\exp(-0.9/T)$ & RRKM \\
 190 & {AlOMg + H$_2$ $\rightarrow$ AlO + Mg + H$_2$} & 326 &4.13(+07)(T/300)$^{-4.23}\exp(-39765.2/T)$ & RRKM \\
 191 & {CaO + Al $\rightarrow$ AlO + Ca} & -84  &3.24(-10)(T/300)$^{-0.20}\exp(6.4/T)$ & RRKM \\
 192 & {AlO + Ca $\rightarrow$ CaO + Al} & 84  &4.99(-10)(T/300)$^{-0.14}\exp(-18766.4/T)$ & RRKM \\
 193 & {AlOH + Ca $\rightarrow$ AlOCa + H} & 114\tablefootnote[1]{calculated at the CCSD/cc-pVTZ level\label{tr1}}  & 1.56(-10)(T/300)$^{2.02}\exp(-13826.3/T)$ & RRKM \\
 194 & {AlOCa + H $\rightarrow$ AlOH + Ca} & -114\textsuperscript{1} &1.03(-12)(T/300)$^{-0.93}\exp(-425.5/T)$ & RRKM \\
 195 & {AlO + Ca + H$_2$ $\rightarrow$ AlOCa + H$_2$}  & -358 \textsuperscript{1} & 1.51(-31)(T/300)$^{-3.42}\exp(-0.9/T)$ & RRKM \\ 
 196 & {AlOCa + H$_2$ $\rightarrow$ AlO + Ca + H$_2$}& 358 \textsuperscript{1} & 3.56(+05)(T/300)$^{-4.28}\exp(-43580.3/T)$ & RRKM \\   
 197 & {Al$_2$O$_2$ + OH + H$_2$ $\rightarrow$ Al$_2$O$_3$H + H$_2$}   & -493  & 4.52(-26)(T/300)$^{-8.63}$& MESMER calculation \\
 198 & {Al$_2$O$_3$H + H$_2$  $\rightarrow$ Al$_2$O$_2$ + OH + H$_2$}   & 493 &6.47(-6)$\exp(-46090/T)$ & MESMER calculation \\
 199 & {Al$_2$O$_3$H + OH $\rightarrow$ Al$_2$O$_4$H + H} & -10 &1.01(-10)(T/300)$^{0.38}$ & MESMER calculation   \\
 200 & {Al$_2$O$_4$H + H $\rightarrow$ Al$_2$O$_3$H + OH} & 10 &1.83(-9)(T/300)$^{0.09}$ & MESMER calculation   \\
 201 & {Mg +   Al$_2$O$_4$H $\rightarrow$ MgAl$_2$O$_4$ + H}   & -11  &3.10(-11)(T/300)$^{-0.19}$ & MESMER calculation   \\
 202 & {MgAl$_2$O$_4$ + H $\rightarrow$ Mg + Al$_2$O$_4$H} & 11 &2.27(-9)(T/300)$^{0.16}$ & MESMER calculation   \\
 203 & {Al$_2$O$_2$ + H$_2$O + H$_2$ $\rightarrow$ Al$_2$O$_3$H$_2$ + H$_2$} &-330  &7.59(-26)(T/300)$^{-9.58}$ & MESMER calculation \\
 204 & {Al$_2$O$_3$H$_2$ +H$_2$ $\rightarrow$ Al$_2$O$_2$ + H$_2$O + H$_2$} & 330  &5.35(-8)$\exp(-23987/T)$ & MESMER calculation \\
 205  & {Al$_2$O$_3$H$_2$ + OH $\rightarrow$ Al$_2$O$_4$H$_2$ + H} &-201  &1.02(-9)(T/300)$^{-0.65}$ & MESMER calculation \\
 206 & {Al$_2$O$_4$H$_2$ + H $\rightarrow$ Al$_2$O$_3$H$_2$ +  OH} & 201 &1.73(-5)(T/300)$^{-1.74}\exp(-23791.4/T)$ & Detailed balance \\
 207 & {Ca + Al$_2$O$_4$H$_2$ $\rightarrow$ CaAl$_2$O$_4$ + H$_2$} & -94  &1.72(-13)(T/300)$\exp(-3755/T)$ & MESMER calculation \\
 208 & {CaAl$_2$O$_4$ + H$_2$ $\rightarrow$ Ca + Al$_2$O$_4$H$_2$} & 94 &4.93(-12)$\exp(-7370/T)$ & MESMER calculation \\
 209 & {Ca + Al$_2$O$_2$ + M $\rightarrow$ CaAl$_2$O$_2$ + M} & -193  &3.06(-28)(T/300)$^{-8.09}$     0.0 & MESMER calculation \\
 210 & {CaAl$_2$O$_2$ + M $\rightarrow$ Ca + Al$_2$O$_2$ + M} & 193  &6.11(-11)$\exp(-20576/T)$ & MESMER calculation \\  
 211 & {CaAl$_2$O$_2$ + OH $\rightarrow$ CaAl$_2$O$_3$ + H} & -270 &1.23(-9)(T/300)$^{0.14}$ & MESMER calculation \\
 212 & {CaAl$_2$O$_3$ + H $\rightarrow$ CaAl$_2$O$_2$ + OH} & 270 &4.98(-7)$\exp(-31331/T)$ & Detailed balance \\
 213 & {CaAl$_2$O$_3$ + OH $\rightarrow$ CaAl$_2$O$_4$ + H} & -179 &8.30(-10)(T/300)$^{-0.70}$ & MESMER calculation \\
 214 & {CaAl$_2$O$_4$ + H $\rightarrow$ CaAl$_2$O$_3$ + OH} & 179 &2.20(-6)$\exp(-19484.1/T)$ & Detailed balance \\
 215 & {CaAl$_2$O$_4$ + SiO + M $\rightarrow$ CaAl$_2$O$_4$SiO + M}  & -492  &2.81(-26)(T/300)$^{-14.59}$ & MESMER calculation  \\
 216 & {CaAl$_2$O$_4$SiO + M $\rightarrow$ CaAl$_2$O$_4$ + SiO +  M} & 492 &2.84(-7)$\exp(-25587/T)$ & MESMER calculation  \\
 217 & {CaAl$_2$O$_4$ + H$_2$O + M $\rightarrow$ CaAl$_2$O$_3$(OH)$_2$ + M} & -318 &2.76(-26)(T/300)$^{-15.06}$  & MESMER calculation  \\
 218 & {CaAl$_2$O$_3$(OH)$_2$ + M $\rightarrow$  CaAl$_2$O4 + H$_2$O + M} & 318 & 5.10(-9)$\exp(-15442/T)$ & MESMER calculation  \\
 219 & {CaAl$_2$O$_4$ + AlO + M $\rightarrow$ CaAl$_3$O5 + M} & -489 &4.05(-24)(T/300)$^{-13.29}$ & MESMER calculation  \\
 220 & {CaAl$_3$O$_5$ + M $\rightarrow$ CaAl$_2$O$_4$ + AlO + M} & 489 &3.17(-5)$\exp(-38785/T)$ & MESMER calculation  \\
\end{tabular}
\end{table*}

\begin{table}
\caption{Cartesian coordinates of the atomic centres of the GM candidate structures presented in this study. 
The full table is available from the CDS via anonymous FTP.\label{coordinates}}
\begin{tabular}{l r r r}
1A \\
O     &    -0.16877  &      0.96174    &    1.72609 \\
O     &     1.18721  &     -0.58226    &   -0.00000 \\
Al    &    -0.16877  &     -0.64481    &   -1.18571 \\
Al    &    -0.16877  &     -0.64481    &    1.18571 \\
O     &    -1.21429  &     -1.42488    &   -0.00000 \\
O    &     -0.16877  &      0.96174    &   -1.72609 \\
Mg   &      0.60876  &      1.45287    &    0.00000 \\
\\
\end{tabular}
\end{table}

\begin{table*}
\caption{Thermochemical tables of the GM cluster candidates. The full table is available from the CDS via anonymous FTP\label{thermotable}.}
\begin{tabular}{c c c c c c c}
1A & \\
 T(K)  &  S (J mol$^{-1}$K$^{-1}$)  & C$_p$ (J mol$^{-1}$K$^{-1}$)  & $\delta\Delta$H (kJ mol$^{-1}$)   &    $\Delta$H$_f$ (kJ mol$^{-1}$) & $\Delta$G$_f$ (kJ mol$^{-1}$) & $\log(K_f)$\\
    0  &     0.000    &       0.000     &      0.000     &  -940.077          & -940.077 & $\infty$ \\
 100   &   255.998    &      47.597     &      3.699     &   -943.794       &  -932.385  & 487.022 \

\end{tabular}
\end{table*}

\begin{table*}
\caption{Fitting parameters a, b, c, d, and e for the computation of $\Delta$G$_{r}^{\circ}(T)$ as defined in \citet{2018MNRAS.479..865S} and used for equilibrium calculations.\label{dgfits}}
\begin{tabular}{r  c c c c c}
Species & a & b & c & d & e \\
\hline
\multicolumn{4}{l}{(MgAl$_2$O$_4$)$_n$} \\
1A             & 3.29349e+05  & -6.69734    & -50.7728     & 0.00400272   & -2.73757e-07 \\
2A             & 7.51831e+05  & -12.1274    &  -131.404    &  0.00828716  & -5.64066e-07 \\
3A             & 1.18824e+06  & -18.555     &  -215.394    &  0.0131985   & -9.01132e-07 \\
4A             & 1.61731e+06  & -24.6412    &  -298.481    &  0.0178873   & -1.22137e-06 \\
5A             & 2.03762e+06  & -30.5908    &  -382.383    &  0.0225431   & -1.54054e-06 \\
6A             & 2.47457e+06  & -37.1498    &  -465.231    &  0.0275476   & -1.88472e-06 \\
7A             & 2.89378e+06 & -42.0406     &  -552.431    &  0.0316631   & -2.16667e-06 \\
\hline
\multicolumn{4}{l}{(CaAl$_2$O$_4$)$_n$}\\
1A             & 3.52652e+05  & -6.38009     & -52.3394     & 0.00395382  & -2.85119e-07 \\
2A             & 7.89531e+05  & -11.6054     & -133.944     & 0.00828386  & -5.94438e-07 \\
3A             & 1.24299e+06  & -17.2971     & -221.608     & 0.012927    & -9.27582e-07 \\
4A             & 1.69168e+06  & -23.2166     & -305.961     & 0.0176701    & -1.26714e-06 \\
5A             & 2.12744e+06  & -28.8081     & -392.233     & 0.0222699    & -1.59754e-06 \\
6A             & 2.57734e+06  & -34.7737     & -478.633     & 0.0270924    & -1.94416e-06 \\
7A             & 3.01948e+06  & -39.6558     & -566.32      & 0.0313433    & -2.25101e-06 \\ 
\hline
\multicolumn{4}{l}{(Mg$_x$Ca$_{(1-x)}$Al$_2$O$_4$)$_n$}\\
2A, x=0.5      & 7.70912e+05 & -11.8584      & -132.685     & 0.00828238   & -5.79068e-07 \\
3A, x=0.67     & 1.20624e+06 & -18.1871      & -217.224     & 0.0131379    & -9.12119e-07 \\
3A, x=0.33     & 1.22616e+06 & -17.8285      & -219.168     & 0.0130755    & -9.22773e-07 \\
4A, x=0.75     & 1.63493e+06 & -24.2286      & -300.652     & 0.0178007    & -1.23048e-06 \\
4A, x=0.5      & 1.65431e+06 & -23.8674      & -302.603     & 0.0177399    & -1.24136e-06 \\
4A, x=0.25    & 1.67139e+06 & -23.4463     & -304.902     & 0.0176569     & -1.25096e-06 \\
5A, x=0.8      & 2.04991e+06 & -30.4413     & -383.478     & 0.0226033     & -1.5601e-06 \\
5A, x=0.6      & 2.06913e+06 & -30.1053     & -385.262     & 0.0225528     & -1.57157e-06 \\
5A, x=0.4      & 2.08729e+06 & -29.6599     & -387.672     & 0.0224543     & -1.58002e-06 \\
5A, x=0.2      & 2.10466e+06 & -29.2119     & -389.906     & 0.0223453     & -1.58738e-06 \\
6A, x=0.5      & 2.52867e+06 & -35.9415     & -471.01      & 0.0272996     & -1.91272e-06 \\
\hline
%
\end{tabular}
\end{table*}


\end{appendix}

\end{document}